\documentclass[12pt]{article}
 
\usepackage[margin=1in]{geometry} 
\usepackage{amsmath,amsthm,amssymb}
\usepackage{MnSymbol}
\usepackage{mathrsfs}
\usepackage{verbatim}
\usepackage{listings}
\usepackage{graphicx}
\usepackage[toc,page]{appendix}
\usepackage{fancybox}
\usepackage{booktabs}
\usepackage{pdfpages}
\usepackage{float}
\usepackage{tikz}
\usetikzlibrary{patterns}
\usepackage{pdflscape}

\RequirePackage[colorlinks,citecolor=blue,urlcolor=blue]{hyperref}

\usepackage{bbm}

\usepackage{listings}
\usepackage{color}
 
\definecolor{codegreen}{rgb}{0,0.6,0}
\definecolor{codegray}{rgb}{0.5,0.5,0.5}
\definecolor{codepurple}{rgb}{0.58,0,0.82}
\definecolor{backcolour}{rgb}{0.95,0.95,0.92}
 
\lstdefinestyle{mystyle}{
    backgroundcolor=\color{backcolour},   
    commentstyle=\color{codegreen},
    keywordstyle=\color{magenta},
    numberstyle=\tiny\color{codegray},
    stringstyle=\color{codepurple},
    basicstyle=\footnotesize,
    breakatwhitespace=false,         
    breaklines=true,                 
    captionpos=b,                    
    keepspaces=true,                 
    numbers=left,                    
    numbersep=5pt,                  
    showspaces=false,                
    showstringspaces=false,
    showtabs=false,                  
    tabsize=2
}
 
\makeatletter
\newcommand*{\rom}[1]{\expandafter\@slowromancap\romannumeral #1@}
\makeatother

\usepackage{subcaption}

\usepackage{natbib}
\usepackage{color}

\begin{document}


\begin{titlepage}

\newcommand{\HRule}{\rule{\linewidth}{0.5mm}} 

\center 
 

\textsc{\LARGE Stevens Institute of Technology}\\[0.5cm] 
\textsc{\Large Financial Engineering}\\[0.5cm] 


\HRule \\[0.4cm]
{ \huge \bfseries  Option Smile Volatility and Implied Probabilities:  Implications of Concavity in IV Curves}\\[0.4cm] 
\HRule \\[0.5cm]
 

\begin{minipage}{0.4\textwidth}
\begin{flushleft} \large
\emph{Author:}\\
Darsh \textsc{Kachhara}\\
John K.E \textsc{Markin}\\
Astha \textsc{Astha}\\
\end{flushleft}
\end{minipage}
~
\begin{minipage}{0.4\textwidth}
\begin{flushright} \large
\emph{Supervisor:} \\
Dr. Dan \textsc{Pirjol} 
\end{flushright}
\end{minipage}\\[2cm]
{\large December 18, 2022}\\[1cm] 
{\large Abstract}\\[0.4cm]
\begin{minipage}{1\textwidth}

Earnings announcements (EADs) are corporate events that provide investors with fundamentally important information. The prospect of stock price rises may also contribute to EADs’ increased volatility. Using data on extremely short-term options, we study that bi-modality in the risk-neutral distribution and concavity in the IV smiles are ubiquitous characteristics before an earnings announcement day. This study compares the returns between concave and non-concave IV smiles to see if the concavity in the IV curve leads to any information about the risk in the market and showcases how investors hedge against extreme volatility during earnings announcements. In fact, our paper shows in the presence of concave IV smiles, investors pay a significant premium to hedge against the uncertainty caused by the forthcoming announcement.\\

\end{minipage}\\[4cm]



{\makebox(0,-440){\includegraphics[scale=0.32]{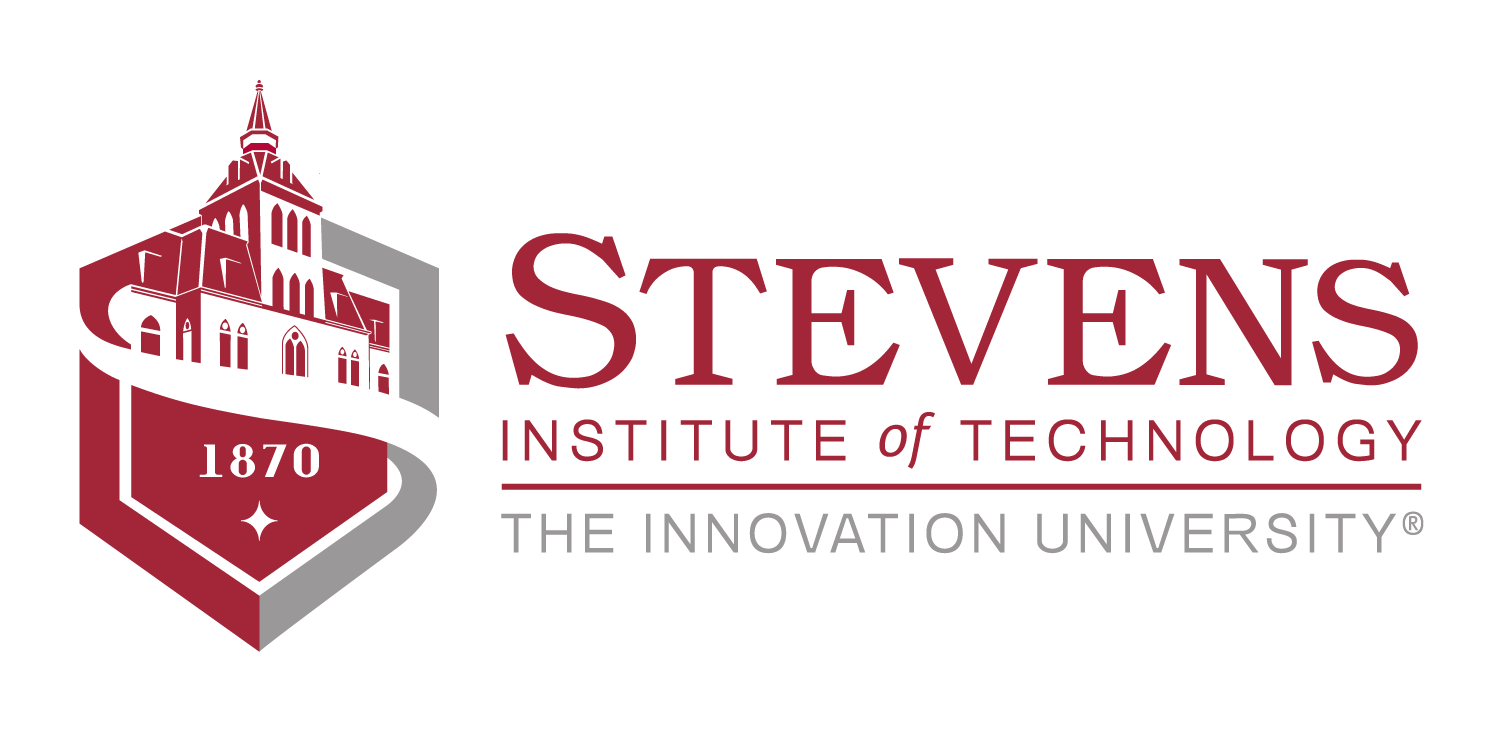}}}%
 

\vfill 

\end{titlepage}


\section{Introduction}
Earnings announcements are corporate events that provide investors with fundamentally important information. Numerous research has examined various facets of the behaviour of stock returns and systematic risk on earnings announcement days (EADs). 
Investors predict that the news will, more often than not, result in a significant change in either direction of the underlying stock price. Due to the anticipated stock price increase, the ex-ante risk-neutral distribution (RND) displays bimodality and the implied volatility (IV) curve exhibits concavity. Using data on extremely short-term options, we establish experimentally that bimodality in the risk-neutral distribution and concavity in the IV curves are ubiquitous characteristics before EADs. The pricing implications of these occurrences are then examined.\\
The prospect of stock price rises may also contribute to EADs’ increased volatility. \cite{patell1981ex} and  \cite{dubinsky2019option} demonstrate an increase in IV before EADs, followed by a sharp fall. In contrast to a more scattered distribution (volatility), a negatively skewed distribution (reflecting tail risk), or a more fat-tailed distribution, the bimodality that we describe is a fundamentally distinct risk notion (kurtosis).\\
The observed bimodality in the core portion of the RND suggests a slight risk adjustment due to the extremely short option expiration. The prevailing stock price is 
anticipated to be around either of the two indicated modes. This suggests that the stock price will likely be x\% higher or y\% lower than the current price following the announcement, with varying levels of volatility accompanying each outcome. We refer to this ex-ante bimodality as an “event risk”\footnote{According to Liu, Longstaff, and Pan (2003, p. 231), event risk is defined as “the risk of a major event precipitating a sudden large shock to security prices and volatilities.”} for the underlying stock and suggest that a concave IV curve signals this risk based on option price.\\
\\
These forms are the inverse of the often observed convex volatility smiles and smirks (or skews) for equities options\citep{hull2003options}, where out-of-the-money (OTM) puts trade at higher volatility than at-the-money (ATM) options. The IV smiles revert to their standard convex shape and the RND loses its bimodality, indicating that the risk associated with the event has vanished, as the concavity feature in our sample disappears almost entirely after the announcement, as the uncertainty surrounding the event is resolved\citep{baker2018trading}.
Our data align with the literature, suggesting concavity is more prevalent in options with shorter expiration dates. The reactive effect of the anticipated jump and the diffusion component of the underlying stock price mechanism causes this feature. 
We integrated two methods to determine the implied volatility numbers. First, we utilise values that are readily accessible on the market. Second, we employ the Fast and Stable approach\citep{jackwerth2004option}, which is a parametric curve fitting method to construct a smooth IV plot for further analysis to determine the implied volatility values for strikes that were not apparent in the market. \\
\\
The concave IV curves we documented are typically inverse W-shaped or inverse U-shaped. This paper also showcases how investors hedge against extreme volatility during earnings announcements. We also compare the returns between a concave and non-concave IV curve to see if the IV curve possesses any information 
about the risk in the market. The most obvious way investors could speculate on or hedge against large stock price swings on EADs, regardless of their direction, is by purchasing straddles. Delta-neutral ATM straddles have been commonly used to capture the price of volatility risk for the underlying stock returns\citep{coval2001expected}.
In fact, we find in the presence of concave IV curves investors pay a significant premium to hedge against the uncertainty caused by the forthcoming announcement. Overall, our study shows that large stock price movements are systematically anticipated by investors before EAD and can be detected ex-ante because they dramatically affect the pricing of short-expiry options. In the case of concave IV curves, we show that large stock price movements are not just a possibility due to the announcement but rather a very likely outcome. This feature gives rise to a bimodal short-term RND for the underlying stock price (and return), which starkly contrasts with the established asset pricing paradigm that relies on unimodal return distributions.

\section{Concave Implied Volatility Curves } \label{sec: IVCurves}
We specifically say that an IV curve is concave if and only if the following circumstances exist. First, given a continuous moneyness range, the second derivative of the fitted Implied volatility curve is negative near the ATM strike. Second, the concavity is halfway between the second-lowest and second-highest strikes of the actual IV points utilized to approximate the smooth IV curve.
The possible worry that the observed concavity could be an artifact of outliers or the used smoothing spline is addressed by these requirements. They ensure our definition does not include highly local infection spots or slenderly concave IV curve regions. Additionally, they ensure that the concavity does not result from the deep OTM options’ deep strikes, which are often the lowest or highest accurate strikes. This concept is broad enough to encompass other concavity curve types, including the inverse U-shape and W-shape IV curves.\\

\begin{figure}[H]
    \centering
 \includegraphics[height=2.5in, width = 3.5in, trim=0 0 0 0, clip]{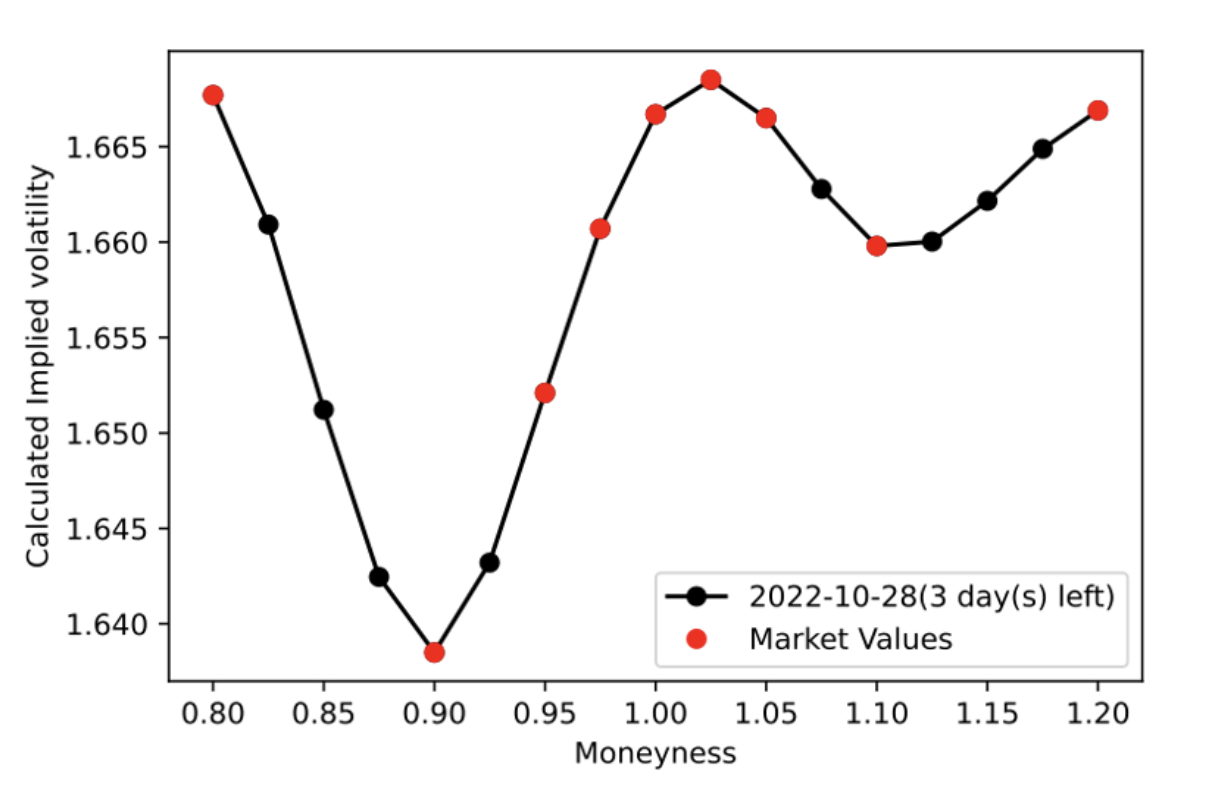}
 \caption{Fig A  W shape concave IV smile}
\end{figure} 

\newpage

\section{Data } \label{sec: Data}
This study’s sample consists of seven corporations (Black Rock, Google, Meta, JP Morgan, Walgreens, Netflix, and Pepsico) analysed across seven quarters beginning in 2021. The data includes implied volatility level (annualised) for the day before, the day of, and the day following the earnings report. This information was obtained from the Bloomberg Terminal dataset BVOL. The data we read from the terminal is based on Bloomberg’s algorithm for calculating the Implied volatility for different strikes. The value is the same for both calls and puts ( which makes comparisons and calculations more straightforward). The dataset contains a mixture of high-growth, high-risk technology corporations that saw strong market tailwinds during the previous year and steady, high dividend-paying equities.
For a more comprehensive conclusion, we analyse the implied volatility levels across three expirations to determine the influence of each expiration. The shortest maturity spans from 1 to 4 days, while the longest extends from 19 to 22 days. We got the announcement time directly from Bloomberg. The sample comprises blue-chip equities to ensure a highly liquid sample.

\section{Methodology } \label{sec:Methodology}

The implied volatility and moneyness of stocks data we take from Bloomberg are not continuous. This makes visualising the smile more difficult because the intervals between any two moneyness points varies. To analyse the bimodality and concavity of a short-maturity options contract, we want smooth smiles that can be built over the collected market data to present a clear image of the phenomena. We employ the Fast and Stable technique for this purpose.
The fast and stable method is theoretically connected to the methods proposed by \cite{jackwerth2004recovering} .This approach is most commonly used to fit the implied volatility smile with a flexible function. The goal is to fit the sums of the squared differences in observed and simulated volatilities. To find the implied volatility for strikes/moneyness that we don’t directly see in the market, we minimize the following objective function:

\begin{equation}\label{eq:jack1}
 min(\sigma_{j})\frac{\Delta^{4}}{2(J+1)}\sum_{j=0}^{j}(\sigma_{j}^{''} )^{2} +  \lambda \sum_{j=0}^{j}(\sigma_{i}  - \sigma_{i}^{-}  )^{2}
    \end{equation}\\

$\sigma_{j} $=implied volatility associated with strike price$ k_{j}$\\
$\Delta $ = difference between two adjacent future index values\\
$ \lambda $= trade-off parameter for balancing smoothness versus fit\\
$\sigma_{j}^{''}$= second derivative of implied volatility concerning strike prices, numerically approximate.\\
$\sigma_{i}$= implied volatility associated with strike price$ k{i}$\\
$\sigma_{j}^{"}$=observed implied volatility of option with strike price $k_{i}$\\

In Appendix \ref{sec:A}, we show that the first-order conditions can be solved in closed form. There are a few advantages to using this curve-fitting algorithm. 
\begin{enumerate}
\item The trade-off between fit and smoothness can be controlled externally,
\item It does not require complicated mathematical functions or optimization routines.
\end{enumerate}
Since we also try to investigate the bimodality in the risk-neutral distributions, we calculate the call prices and numerically approximate the distribution. The probability density function is the second derivative of the call prices with respect to K, i.e. $G(k)= e^{rT}(d^{2}c/dk^{2})$, where r is the risk-free rate. The result, which is from \cite{breeden1978prices}, allows risk-neutral probability distributions to be estimated from volatility smiles. Suppose that c1, c2, and c3 are the prices of the equity call options with strike prices of K-, K, and K+, respectively. Assuming it is small, an estimate of G(k) is obtained by approximating the partial derivative as  

  \begin{equation}\label{eq:gk}
G(k)\approx e^{rt}(c_{1} + c_{3} - 2c_{2}) /\Delta^{2}
    \end{equation}\\

Now, after calculating, we see that these are not probabilities because they do not sum to 1, so we standardise them by dividing each by the sum of all values. We have inputted nine values from the market ( the one we observed ) for our sample and returned 17 values using the curve fitting algorithm. The moneyness/strikes for our volatility smile are 0.025 points away from each other. We calculate the Implied volatility and Risk neutral distributions for three expiries to showcase that the effect of extreme risk is visible most of the time in the shortest maturity. To calculate the volatilities, we set trade-off parameters empirically so that we price the options accurately enough for our purposes. We reduce $\lambda$ until all risk-neutral probability distributions are nonnegative—thus, arbitrage-free, we take our $\lambda$ value to 0.01\footnote{See Appendix A} . The one caveat we need to mention here is that the Fast and Stable method only works for European options. This would’ve created a problem for us since we are using American.options, but the Bloomberg values we use as inputs are calculated in a way that they represent the European options \footnote{Bloomberg, “Real-Time Volatilities a Service to Access Implied Market Volatilities, accessed December 18, 2022, https//assets.bbhub.io/professional/sites/10/978935Real-Time-Volatilities.pdf."}

\section{Features of Concave Implied Volatility Curves and Bimodal Risk-Neutral Distribution } \label{sec:Features}
To understand the concavity impact in implied volatility smiles and the resulting bi-modality in risk-neutral distribution, we examine the last seven quarters of seven companies beginning in 2021. 40\% of our sample had concave IV smiles before the earnings release. Surprisingly, we also saw concave IV smiles on the day of EAD, which demonstrated market participants’ indecision and lack of confidence regarding the prognosis of a specific stock’s price, even after the information that was the cause of the risk had been disclosed. This led to the delta and gamma hedge by the participants by the participants \cite{alexiou2021pricing}which kept the concave smile intact. Another phenomenon we witness when we examine the concave IV smile is an instant drop in implied volatility levels one day after the announcement (in the shortest maturity traded, changing the IV smile shape back to convex).

\subsection{ Study on Concave IV: BlackRock Inc (BLK)  } \label{sec:BLK}

\begin{figure}[H]
    \centering
 \includegraphics[height=2.7in, width = 5.7in, trim=0 0 0 0, clip]{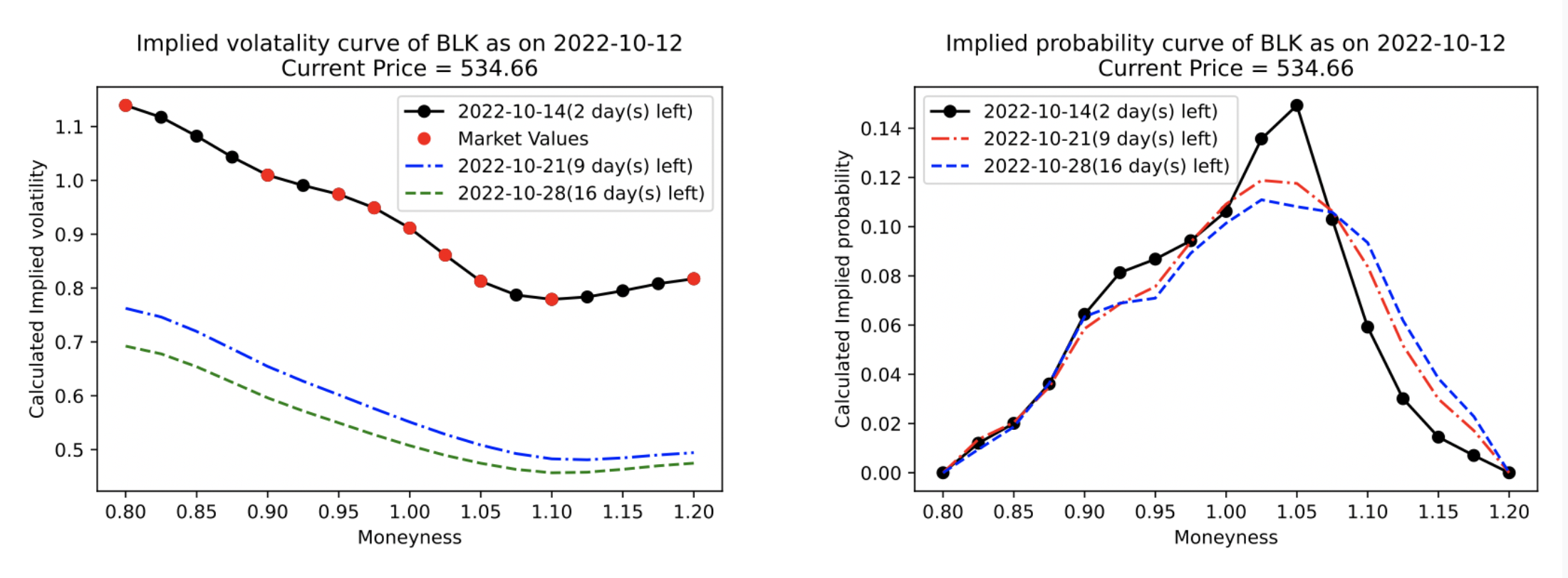}
 \caption{BLK Q3 2022 ( Fig 1 ) }
\end{figure} 
Black Rock has been going through a rough patch since Q1 2021. The stock has returned -21.7\% from the beginning of 2021 till 18th October 2022. The Price to Earnings ratio in the same time period was down by -33\%. For  Q3 2022, the predicted Earnings per share was \$7.032. The market was predicting lower earnings per share than last quarter owing to the downward trend in company performance. But, in this particular quarter, there seemed to be a  disagreement regarding this estimate. Hence, we see concavity and b-imodality in the IV smile and Risk neutral distribution, respectively, on the day before the earnings announcement. 
The reported Earnings per share was \$9.25 which is 31\% more than the estimation. This also led to a “jump” in price as the stock returned 5.6\% after the announcement. The concavity we study in this quarter is known as the “W” shape concavity. This is what the IV smile looked like one day after the announcement,\label{sec:BLK2}

\begin{figure}[H]
    \centering
 \includegraphics[height=2.7in, width = 5.7in, trim=0 0 0 0, clip]{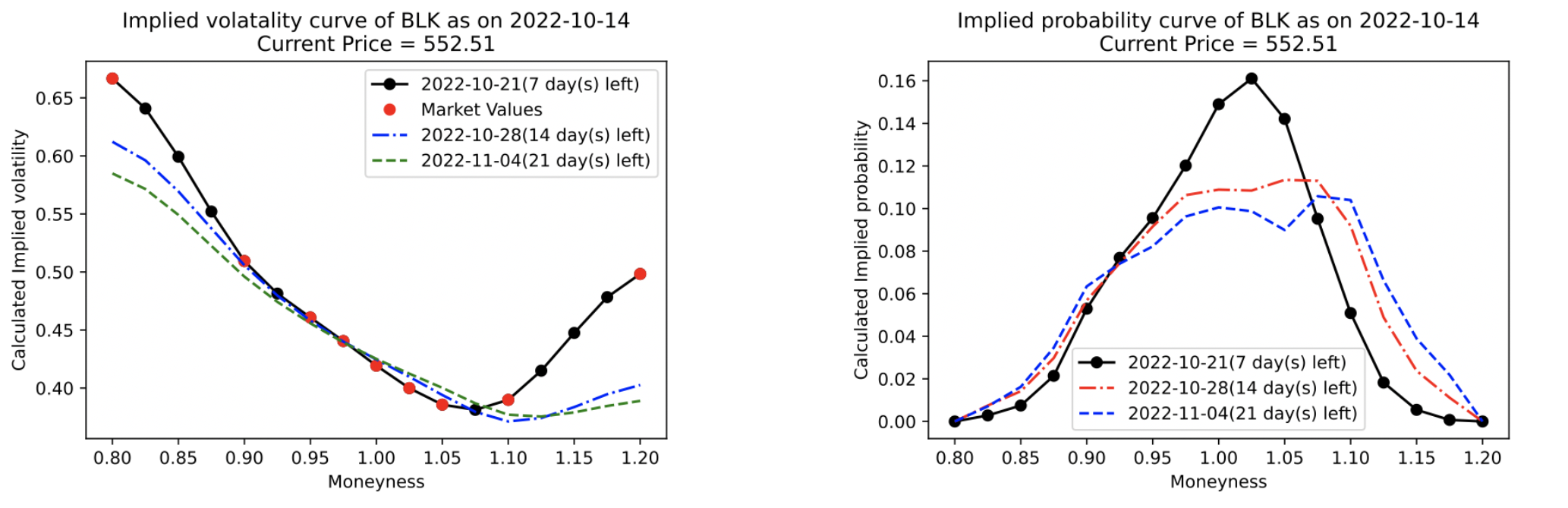}
 \caption{BLK Q3 2022 (Fig 1A). }
\end{figure} 

The stock corrected itself one day after the announcement yet gave a positive outcome overall. As you can see, the immediate expiry has 7 days left because the previous contract expired on the announcement day. But, we see the IV smile for the next immediate expiry has returned to its convex shape, although there is still some uncertainty in the longer maturities.\\

\subsection{ Study on Concave IV: Alphabet Inc (GOOGL) } \label{sec:googl}

\begin{table}[H]
\caption{GOOGL table 1}
\label{tab:TableOptionCharac1}
\centering
\begin{tabular}{ccc}
  \toprule
   Quarter & Shape& Return of stock  \\
  \midrule 
  Q4 2021 & “W” shape	 & 9.89\% \\
  \midrule
  Q1 2022& “W” shape& -7.50\%\\
  \midrule
 Q2 2022 (ON EAD)& “W” shape & -1.85\% \\
  \midrule
  Q2 2022 & “W” shape& 6.16\% \\
  \midrule
  Q3 2022&“W” shape & -7.61\% \\
  \bottomrule
\end{tabular}
\end{table}

For Alphabet, we witness a concave IV smile in 4 quarters out of 7. One of the reasons for the high number of concavity is that the volatility in the technological sector in the past year has been high vis-a-vis the market\footnote{Tim Culpan, “Tech Stocks: 10 Charts Explain the Industry's Challenges,” Bloomberg.com, https://www.bloomberg.com/opinion/articles/2022-10-29/tech-stocks-10-charts-explain-the-industry-s-challenges.}. Also, the increase in the Fed-funds rate increases the risk-free rate of interest, which majorly affects the valuations of technological companies\footnote{Jeran Wittenstein, “Tech-Heavy Nasdaq 100 Index Is at Fed's Whim with Rate Hike Looming,” Bloomberg.com,https://www.bloomberg.com/news/articles/2022-03-16/tech-stocks-are-at-fed-s-whim-with-rate-hike-looming-tech-watch.}. Alphabet, a behemoth in this industry, would be sensitive to these risks. The quarter where the concave IV smile and bimodal risk-neutral distribution were the most prominent was Q2 2022. 

\begin{figure}[H]
    \centering
 \includegraphics[height=2.7in, width = 5.7in, trim=0 0 0 0, clip]{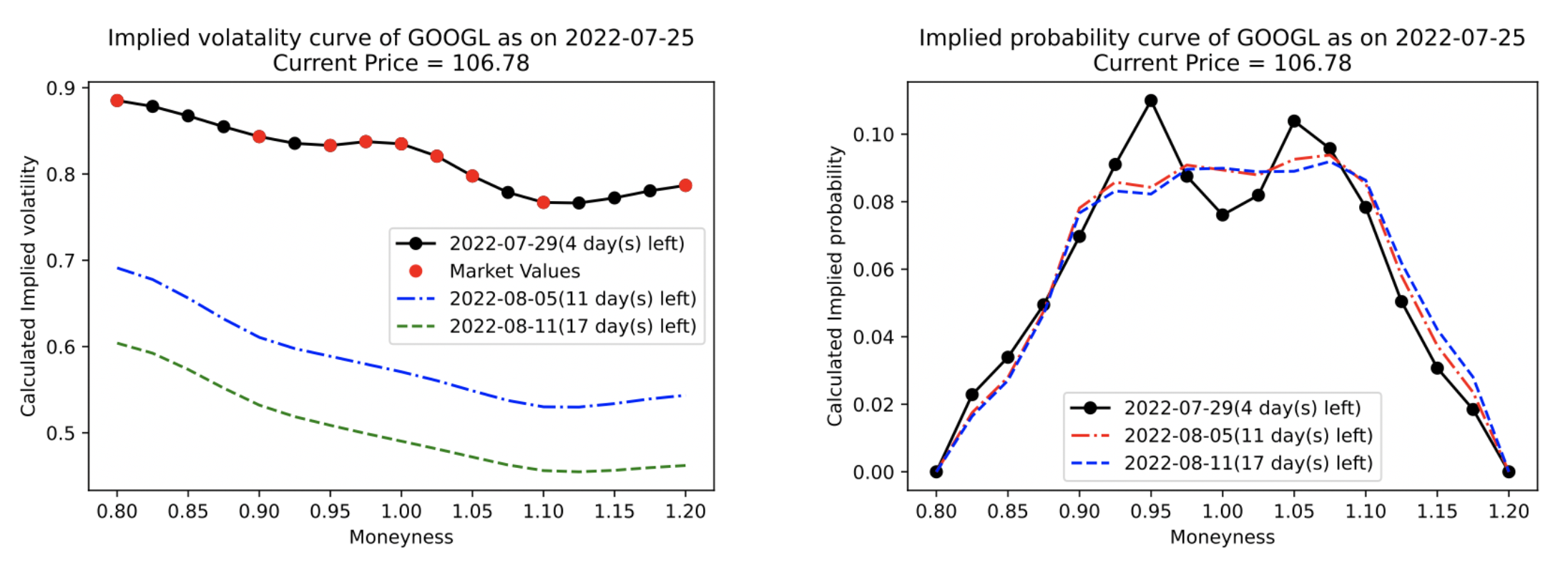}
 \caption{GOOGL Q2 2022 (Fig 2). }
\end{figure}

The Alphabet stock has returned 22.63\% until Q2 2022 with an annualised realised volatility of almost 30\%. The Earnings per share projected for this quarter was \$1.38, 10\% lower than the last quarter. Fig 2 shows the “W” shape concavity in the IV smile and the bimodality in the implied risk-neutral distribution. What’s surprising about this quarter is that although the reported Earnings per share was 12.32\% lower than the projected at \$1.21, the stock still returned 6\% one day after the announcement. The reason for this jump in the stock price is unknown but what’s important to note is that the IV smile and the bimodality indicated show one of the probable outcomes to be a higher stock price in the shortest maturity options contract one day before the earnings announcement. The uncertainty continued even on the close of the day of the announcement. \\

\begin{figure}[H]
    \centering
 \includegraphics[height=2.7in, width = 5.7in, trim=0 0 0 0, clip]{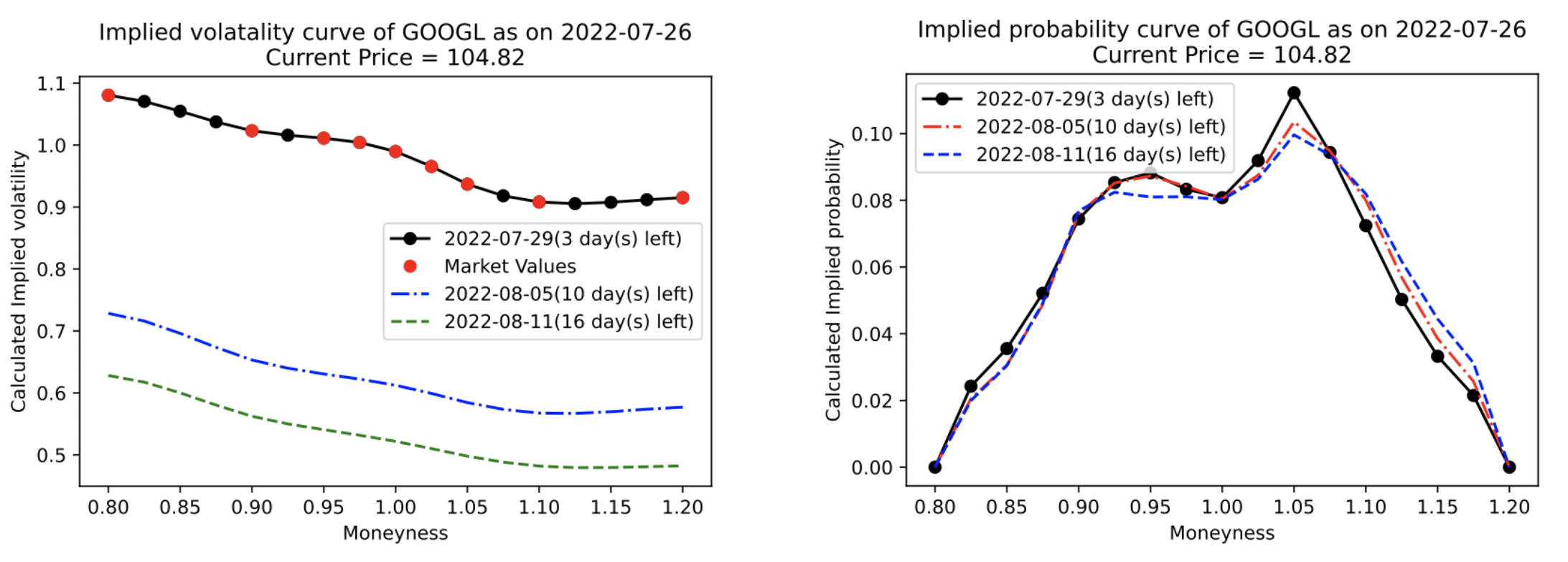}
 \caption{GOOGL Q2 2022 (Fig 2A).}
\end{figure}

\begin{figure}[H]
    \centering
 \includegraphics[height=2.7in, width = 5.7in, trim=0 0 0 0, clip]{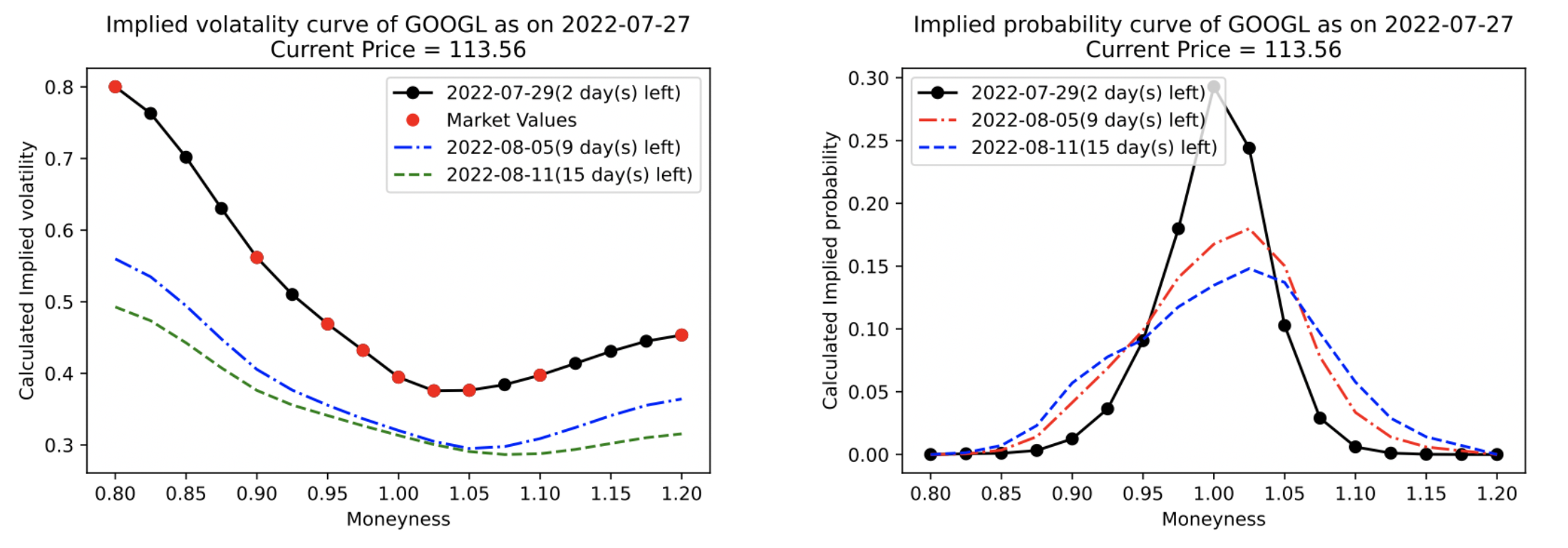}
 \caption{GOOGL Q2 2022 (Fig 2B). IV 1 Day after the Announcement} 
\end{figure}

\subsection{ Study on Concave IV: Meta Inc  (META) } \label{sec:googl}

\begin{table}[H]
\caption{META Table 1}
\label{tab:TableOptionCharac1}
\centering
\begin{tabular}{ccc}
  \toprule
   Quarter & Shape& Return of stock  \\
  \midrule 
  Q1 2021 & “W” shape	 & 8.09\% \\
  \midrule
  Q4 2021 & “W” shape& --29.54\%\\
  \midrule
 Q1 2022 & “W” shape & -12.33\% \\
  \midrule
  Q2 2022 & “W” shape& 0.49\% (6\% jump on EAD) \\
  \midrule
Q3 2022 ( on EAD)& “Inverse U"& -5.93\%  \\
  \midrule
Q3 2022 & “W” shape& -33.89\%  \\

  \bottomrule
\end{tabular}
\end{table}

In Meta, we witness a concave IV smile in 5 quarters out of 7. Meta has been going through a risky shift \footnote{Mac, Ryan, Sheera Frenkel, and Kevin Roose. “Skepticism, Confusion, Frustration: Inside Mark Zuckerberg's Metaverse Struggles.” The New York Times. The New York Times, October 10, 2022. https://www.nytimes.com/2022/10/09/technology/meta-zuckerberg-metaverse.html. }regarding its business operations and primary product. A social media giant aspires to be a metaverse, with virtual reality/augmentation being its primary technological development. The hardware and software required for this transition come at a cost, not to mention the cost of acquiring talent. The stock has dropped by about -62\% in value since the beginning of 2021 (till Q3 2022) and the price-to-earnings ratio for the same time period has declined by over -50\%. This is in line with the current negative sentiment of the company due to the addictiveness of its applications, failure to address the misinformation problem on its platform, data leaks, legal troubles and deep investments into re-building the image and goal of the company\footnote{ Ananya Bhattacharya, “Facebook's Name Change Is Not Enough to Fix Its Reputation,” Quartz (Quartz, October 26, 2021), https://qz.com/2077531/can-a-name-change-fix-facebooks-reputation.}. To visually see these risks and uncertainty being reflected in the market, we look at the concave IV smile of Q3 2022. This quarter also witnessed the highest negative jump, where prices fell by -33\% one day after the announcement. The negative jump was anticipated by the options market seeing the “W” shape concave IV smile one day before the announcement and the inverse U shape smile on the day of the announcement.\\

\begin{figure}[H]
    \centering
 \includegraphics[height=2.7in, width = 5.7in, trim=0 0 0 0, clip]{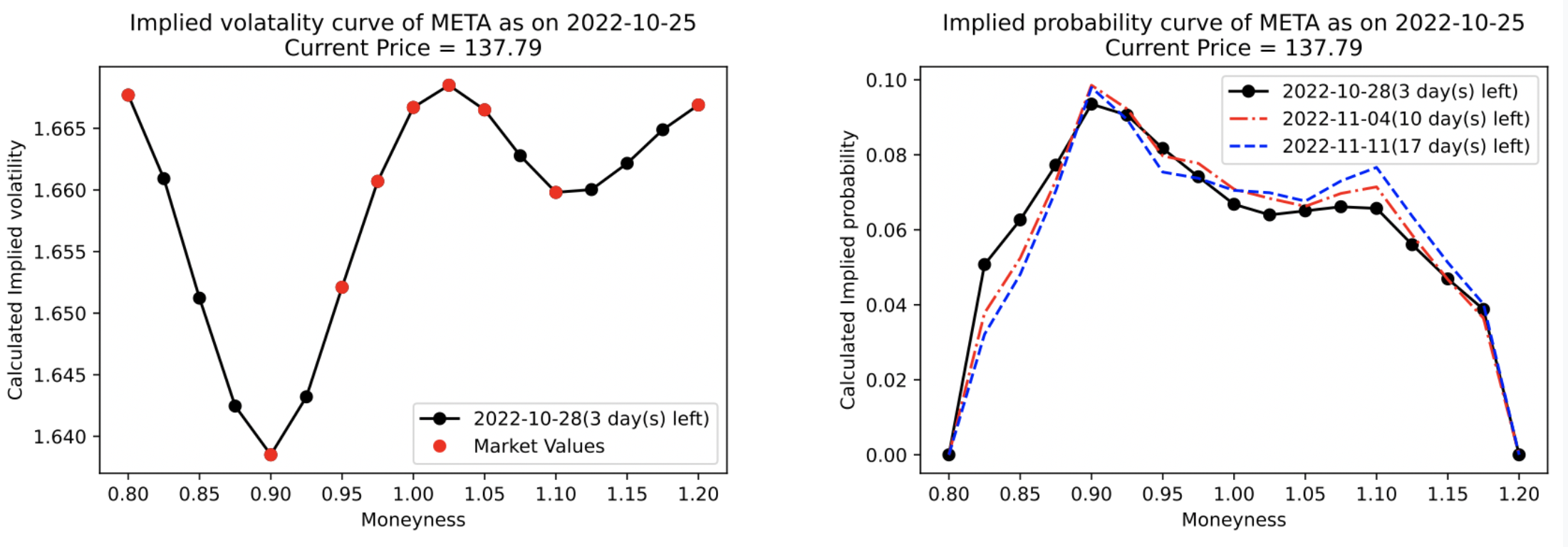}
 \caption{META Q3 2022 (Fig 3) .see note 8} 
\end{figure} 
 \footnote{The images of IV smile where we only display the shortest maturity - This is due to the fact that IV levels for the shortest maturity were quite high compared to the longer maturity. As a result, the shape of the smile was getting obscured if all three curves were shown in the same plot.}\\
 
 The estimated earnings per share for Q3 2022 were \$1.885, whereas the reported earnings per share were \$1.64, -13\% less than expected. If we look at the volatility, the annualised realised volatility till the quarter was 49\%, whereas the annualised ATM implied volatility just before the announcement was more than 165\%. The bimodality in risk-neutral distribution provides a higher probability of a negative outcome than a positive outcome. On the earnings announcement day, the stock faced a drawdown of -5.93\%. However, the concavity and bimodality in the IV smile and risk-neutral distributions were still visible. Here, we notice an inverse U shape concave IV smile with ATM annualised implied volatility exceeding 200\%.\\

\begin{figure}[H]
    \centering
 \includegraphics[height=2.7in, width = 5.7in, trim=0 0 0 0, clip]{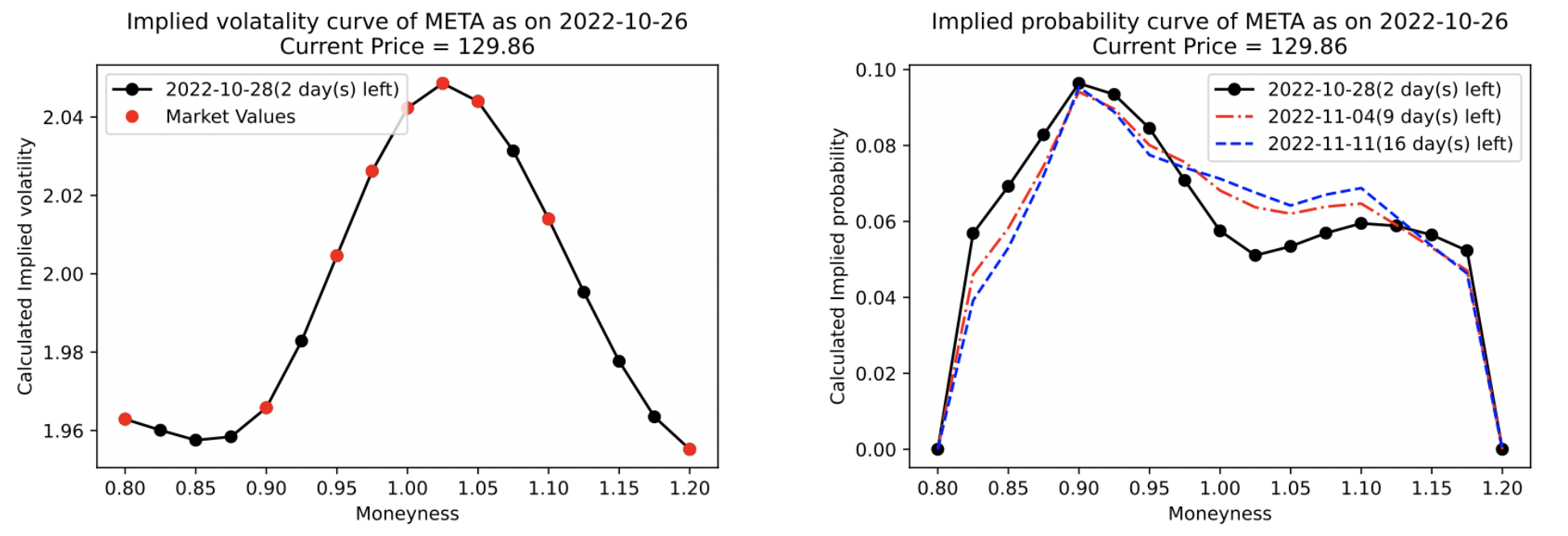}
 \caption{META Q3 2022 (Fig 3A). see note 8} 
\end{figure} 

The stock dropped by -33.89\% one day after the announcement. Fig 3B was the IV smile one day after the announcement. The annualised ATM implied volatility was 87.5\%. Although the implied risk-neutral curve changed its shape to unimodal, the longer expires are still expecting an uncertain outcome.\\

\begin{figure}[H]
    \centering
 \includegraphics[height=2.7in, width = 5.7in, trim=0 0 0 0, clip]{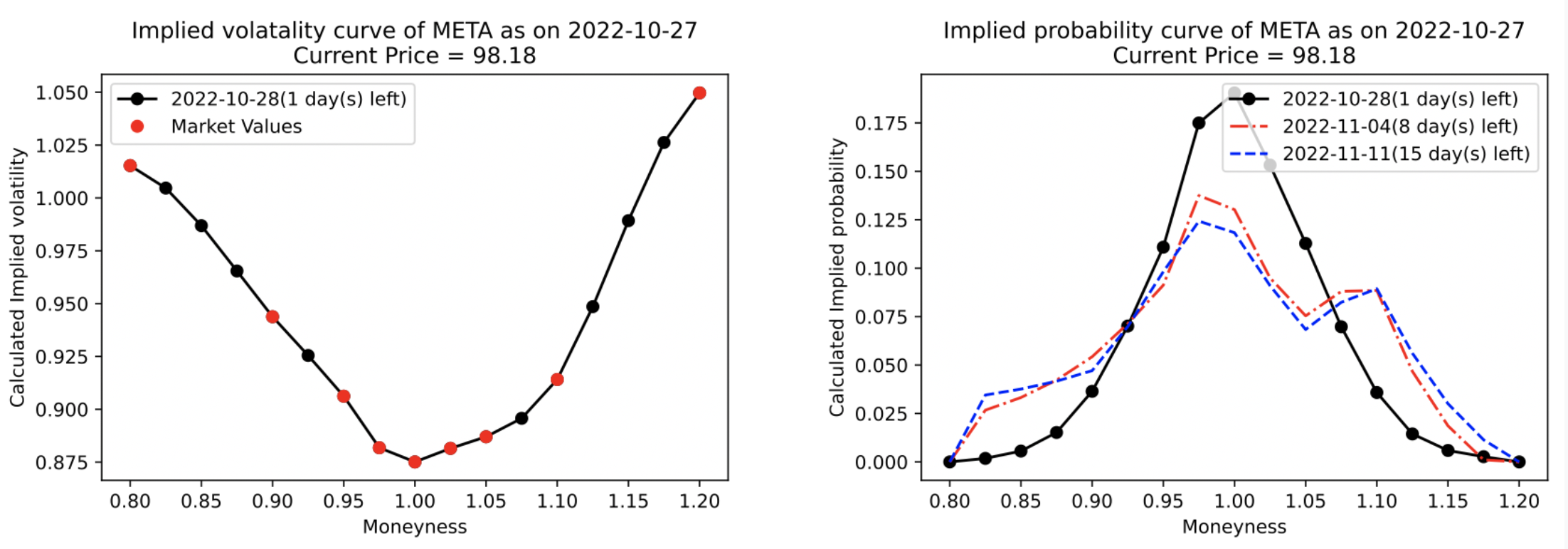}
 \caption{META Q3 2022 (Fig 3B). see note 8} 
\end{figure} 

Something strange that we examined that wasn’t mentioned in the literature was a “Trimodal” risk-neutral distribution. This distribution was visible in Q1 2022. Fig 3C shows the before and after earnings announcement IV smile and Risk Neutral distribution. Fig 3C shows the before and after of the Q1 2022 earnings announcement.\\
\begin{figure}[H]
    \centering
 \includegraphics[height=5in, width = 5.7in, trim=0 0 0 0, clip]{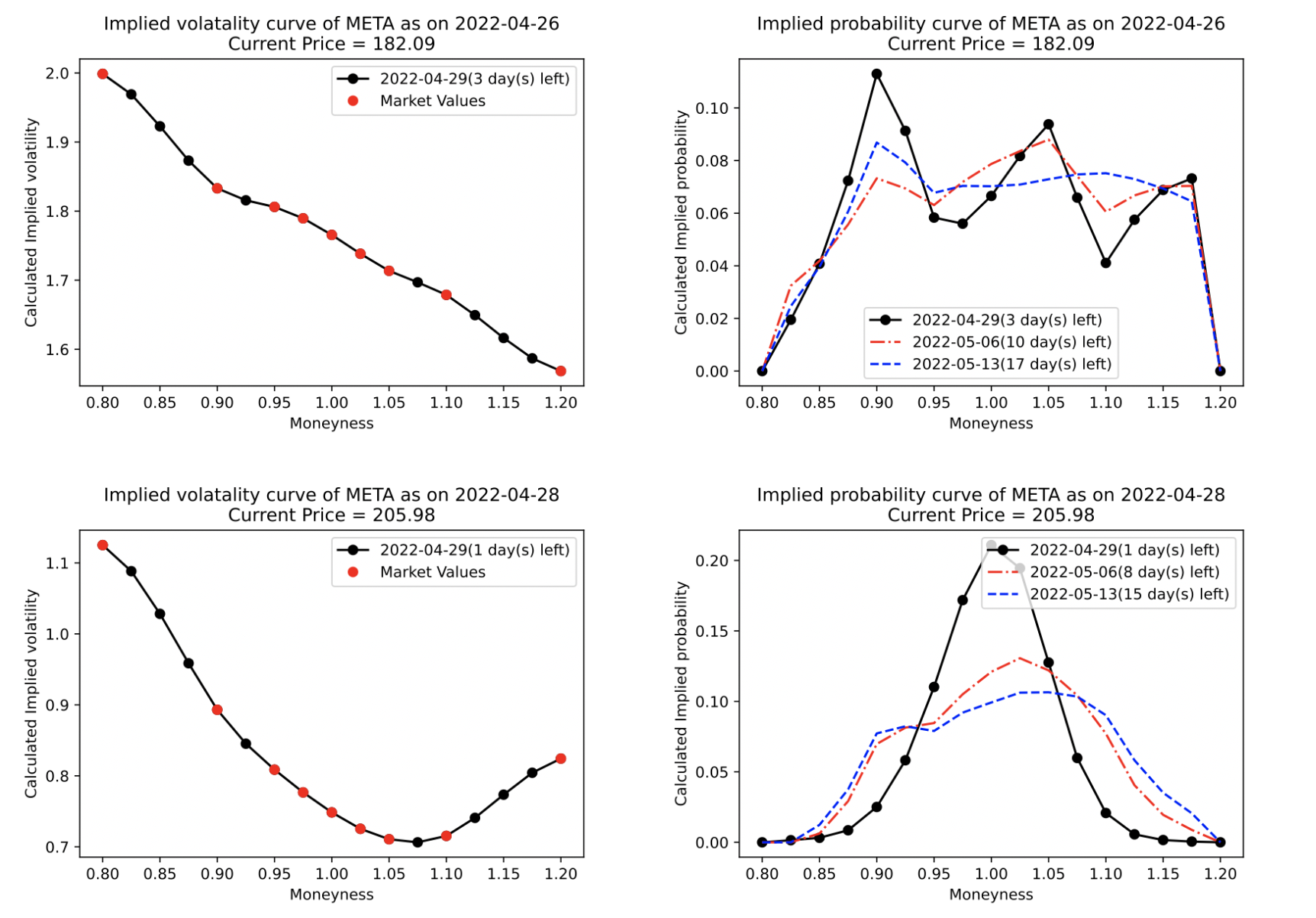}
 \caption{META Q1 2022 (Fig 3C). see note 8} 
\end{figure}

\subsection{ Study on Concave IV: Neflix Inc (NFLX) } \label{sec:nflx}

\begin{table}[H]
\caption{NFLX Table 1}
\label{tab:TableOptionCharac1}
\centering
\begin{tabular}{ccc}
  \toprule
   Quarter & Shape& Return of stock  \\
  \midrule 
  2021Q1 & “W” shape	 & -8.55\% \\
  \midrule
 2021Q4 & “W” shape& -3.62\%\\
  \midrule
 2022Q2 & “W” shape & -27.97\% \\
  \midrule
  2022Q3& “W” shape& -40.93\%\\
  \midrule
Q3 2022 ( on EAD)& “Inverse U"& 12.35\%  \\
  \midrule
Q3 2022 & “W” shape& 9.04\%  \\
  \bottomrule
\end{tabular}
\end{table}

While examining Netflix, we came across 6 concave IV smiles out of 7. Most of these concave smiles were followed by significant jumps ( or drawdowns)  in the stock price. Netflix faced fierce competition\footnote{sherman4949, “Netflix Quietly Admits Streaming Competition Is Eating into Growth,” CNBC (CNBC, January 20, 2022), https://www.cnbc.com/2022/01/20/netflix-quietly-admits-streaming-competition-is-eating-into-growth.html.} from platforms such as  Disney+, HBO max and Amazon prime. Being a growth stock, Netflix is expected to add consumers to its platform, to justify the heavy investments in content building and sophisticated algorithms that make the user experience unique on its platform. The massive drawdowns after the Q4 2021 announcement and Q1 2022 announcement were brought on by the loss in subscribers worldwide.  Netflix announced in early 2022 that it expects to lose close to 2 million subscribers\footnote{Nicole Sperling, “Netflix Expects to Lose 2 Million Subscribers by July,” The Seattle Times (The Seattle Times Company, April 19, 2022), https://www.seattletimes.com/business/netflix-expects-to-lose-2-million-subscribers-by-july/.}.\\
\begin{figure}[H]
    \centering
 \includegraphics[height=5in, width = 5.7in, trim=0 0 0 0, clip]{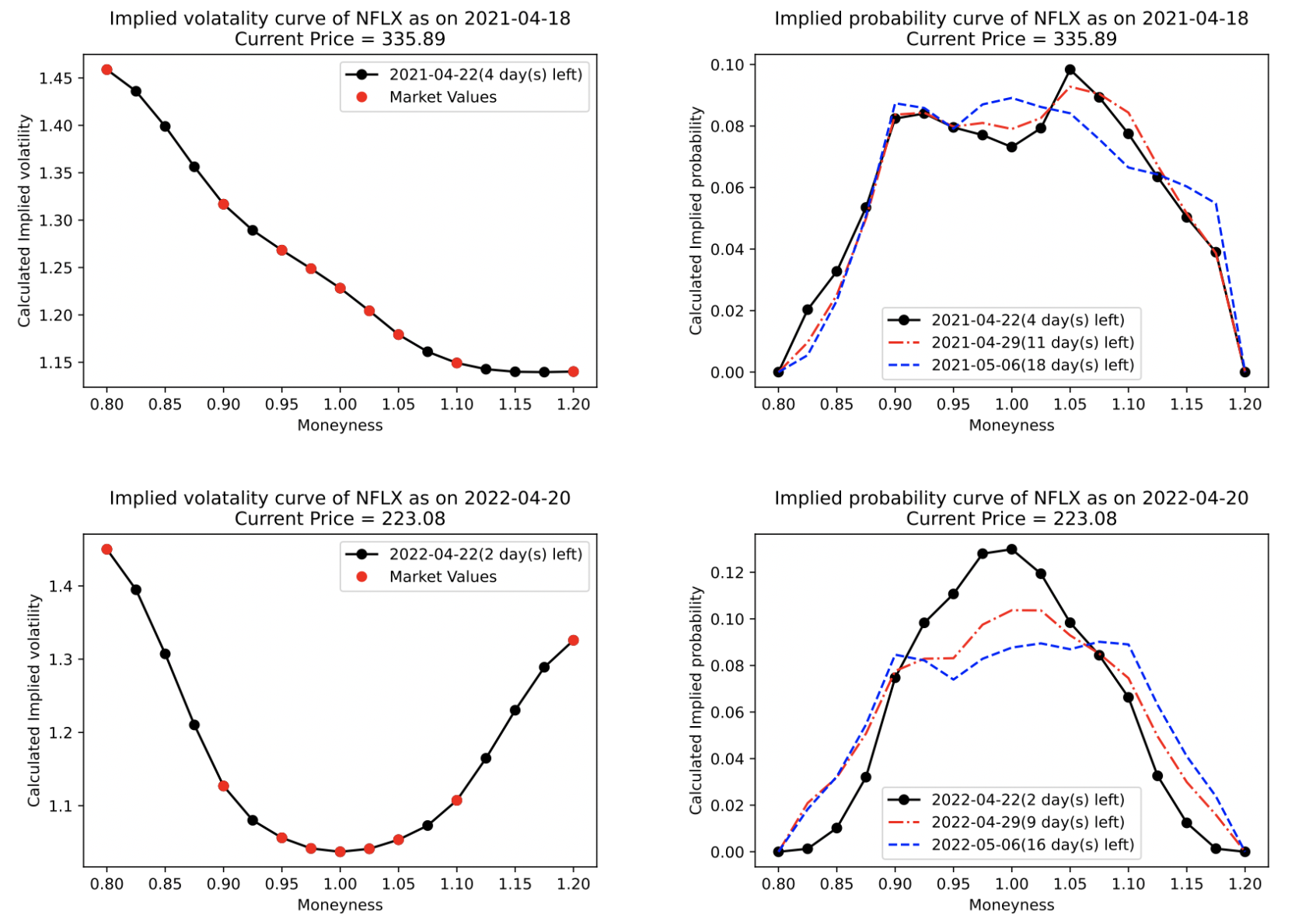}
 \caption{NFLX Q1 2022 (Fig 4). see note 9} 
 \label{fig:nflx1}
\end{figure} 

Figure \ref{fig:nflx1} shows the before and after earnings announcement IV smile and risk-neutral distribution. For our paper, we will also discuss Q2 2022 as the shape of the concave IV smile is more prominently visible, than it is in other quarters.\\

\begin{figure}[H]
    \centering
\includegraphics[height=2.7in, width = 5.7in, trim=0 0 0 0, clip]{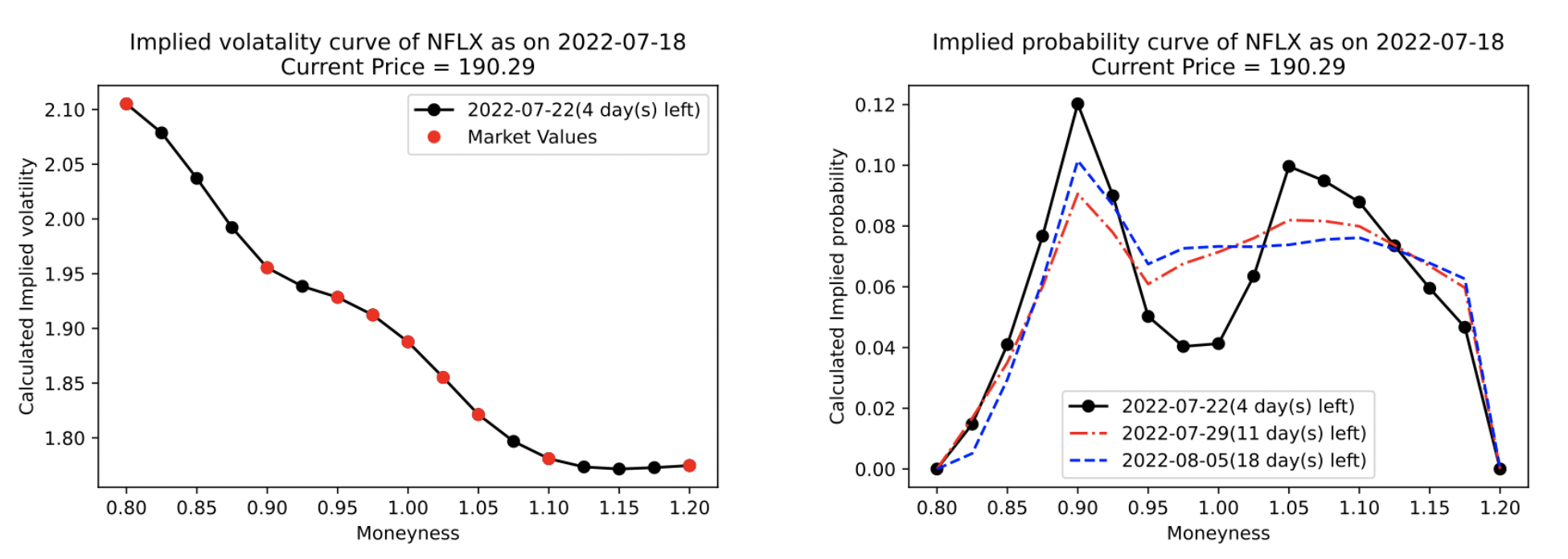}
 \caption{NFLX Q2 2022 (Fig 4A). see note 8} 
\end{figure} 

In contrast to prior quarters, the stock returned a 12.35\% increase following the announcement for this quarter. Netflix revealed financial figures for the second quarter of the fiscal year 2022 that exceeded analysts’ estimates. Netflix reported earnings per share (EPS) of \$3.20, a 7.7\% rise year over year (YOY) and more than the \$2.93 projected by analysts. This represents a dramatic turnaround of Netflix’s YOY EPS drop in Q1 FY 2022. Revenue increased by 8.6\% year on year, matching analyst expectations. Netflix’s income increased at its slowest rate in previous years. Figure 4B depicts the IV smile and risk-neutral distribution following the announcement.\\

\begin{figure}[H]
    \centering
\includegraphics[height=2.7in, width = 5.7in, trim=0 0 0 0, clip]{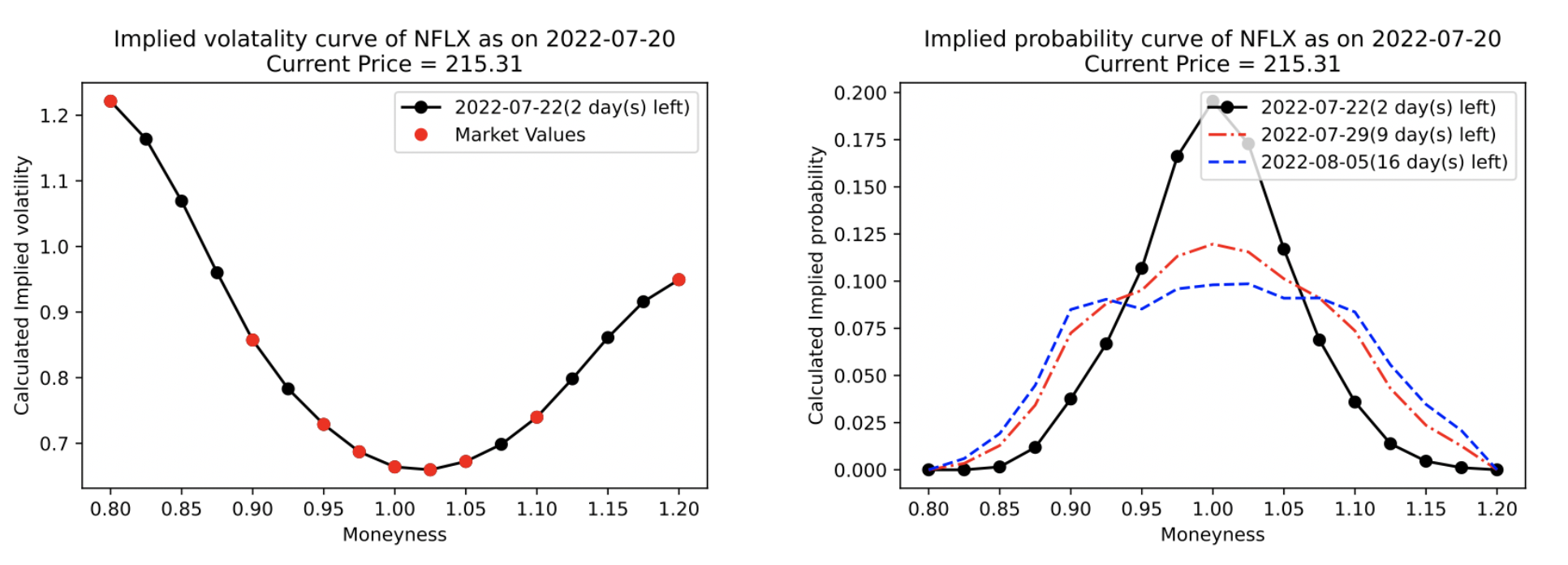}
 \caption{NFLX Q2 2022 (Fig 4B). see note 8} 
\end{figure}

\subsection{ Study on Concave IV: Walgreens Boots Alliance Inc (WBA) } \label{sec:WBA}

\begin{table}[H]
\caption{WBA Table 1}
\label{tab:TableOptionCharac1}
\centering
\begin{tabular}{ccc}
  \toprule
   Quarter & Shape& Return of stock  \\
  \midrule 
  2021Q3 & “W” shape	 & -8.58\% \\
  \midrule
 2022Q2 & “W” shape& -7.88\%\\
  \midrule
 2022Q4 & “W” shape & 3.91\% \\
  \bottomrule
\end{tabular}
\end{table}

Walgreens Boots Alliance corporation has been facing stiff competition from CVS. The pharmaceutical giant has been witnessing slow growth in revenue and some unimpressive quarters. The stock price has lost (till Q4 2022)  -29\% value since the beginning of 2021. The concavity in IV smile was most prominent in Q4 2022. \\

\begin{figure}[H]
    \centering
\includegraphics[height=2.7in, width = 5.7in, trim=0 0 0 0, clip]{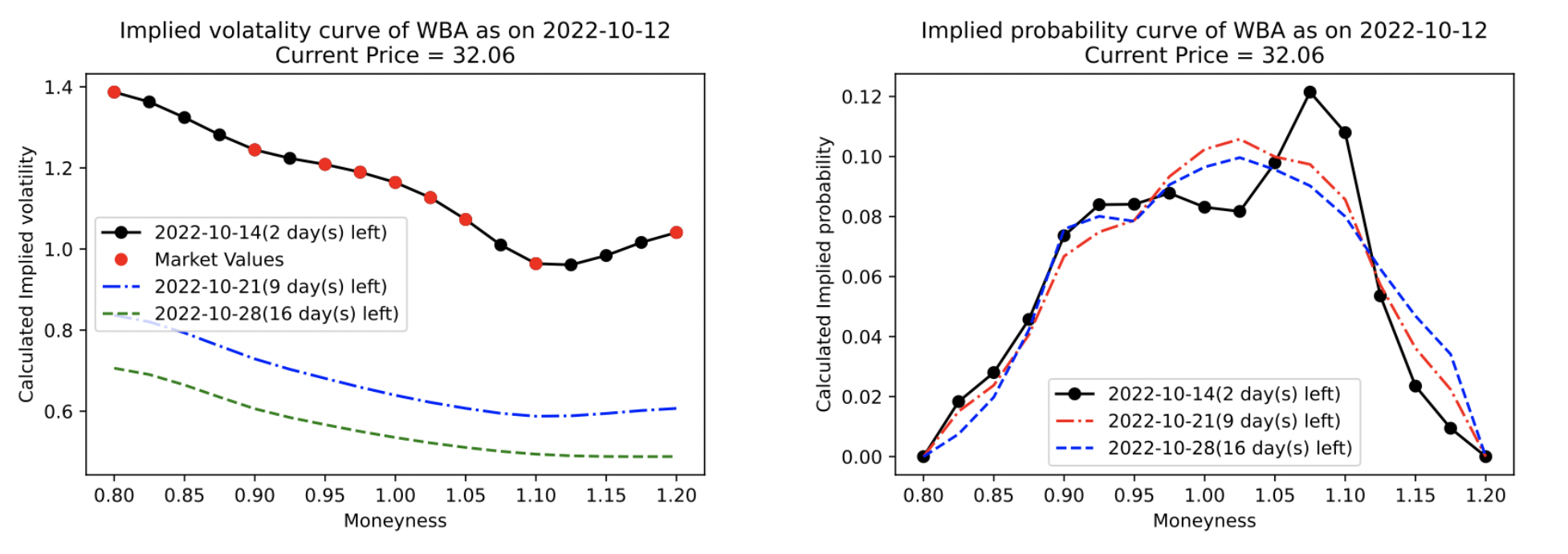}
 \caption{WBA Q4 2022 (Fig 5)} 
\end{figure}

This particular quarter was comparatively better for Wallgreens; their reported earnings per share were in line with the street’s expectations. But, there are still anticipations of lower-than-estimated earnings per share due to the strong dollar and decrease in covid vaccine demand. The company announced that it expected a positive outlook in 2023\footnote{melissa repko, “Walgreens Beats Sales Expectations, as It Expands Its Health-Care Business,” CNBC (CNBC, October 13, 2022), https://www.cnbc.com/2022/10/13/walgreens-wba-q4-2022-earnings.html.}. After the announcement, the respective curves returned to their ‘normal’ shapes.\\
Unfortunately, we did not find concavity in JP Morgan (JPM) and Pepsico (PEP). Since our sample was chosen at random, this occurrence was expected. These companies are well-established, cash-rich and widely analysed; it’s possible that the market was never uncertain of the risky outcomes to a point where concavity would be visible. We further explain the implications of concavity in IV smile and bi-modality in risk-neutral distribution using options hedging strategies.\\

\begin{figure}[H]
    \centering
\includegraphics[height=2.7in, width = 5.7in, trim=0 0 0 0, clip]{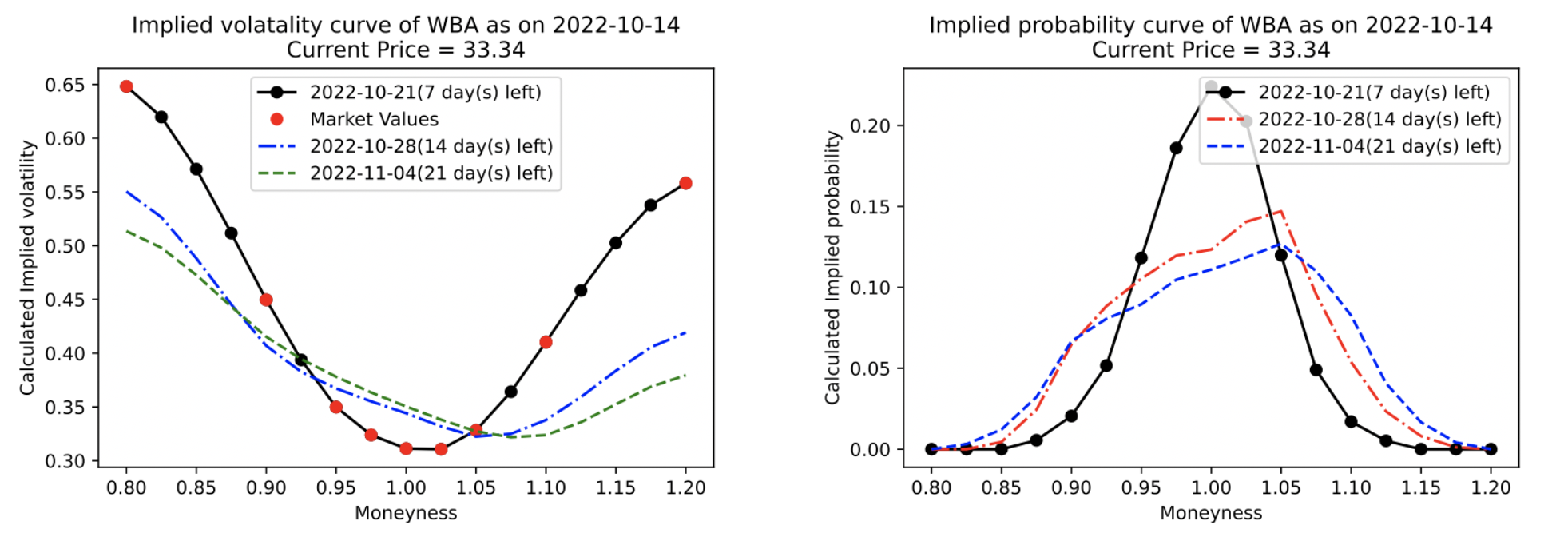}
 \caption{WBA Q4 2022 (Fig 5A)} 
\end{figure} 
This concludes our sample space of concave IV smiles and bimodal risk-neutral distribution. Our study was in line with the literature; concavity in the smiles was accompanied by bi-modality in the distributions. There were visible when risk was anticipated in the market and the participants were uncertain about their expectations and outlook of the company. The IV smile was almost always followed by a sharp jump/ drawdown in the value of the stock. \\

\section{Implications of Concave IV Curves } \label{sec:Implications}
As the implied volatility for an option increases the probability of a positive payoff increases and hence the option is more valuable. However, the direction in which the volatility moves is not easily quantified but subject to market reaction. Examining concave implied volatility curves (IV) before earnings announcement, an event where the implied volatility curve exhibits large fluctuations in the forms of a concave bell can forecast the uncertainty about price direction. The information content in concave IV curves provides a signal for ex-ante predictions for higher or lower absolute stock returns. Evidence from \cite{beaver1968information}, \cite{kapadia2019idiosyncratic} and \cite{banerjee2013abnormal} imply stock prices often exhibit very large movements around earnings. In \cite{alexiou2021pricing} the study computed the abnormal stock return on EAD as the realized minus the expected return, the results showed a significant positive return on average for concave smiles on earnings announcement than without. An interpretation of the predictive information then is that concave IV curves provide an ex-ante forecast to identify where earnings announcements where larger than average stock price movements are observed \cite{alexiou2021pricing} . This paper will investigate the returns on earning announcement day for concave and non-concave smiles assuming there is no cost to trade these assets, isolating the magnitude of the returns.

\subsection{Formation of Straddles and Implications on IV Curve}
When implied volatility curves illustrate concavity on earning announcements, the underlying motivation is that market participants are uncertain of the future direction of price movements due to informational content to be revealed in the announcement,
This uncertainty is viewed in the risk neutral distribution, a hump represents the directional stance of positive news against negative news this shift the current price in either direction. Long position on an at the money (ATM) call and a put generates a payoff structure that neutralizes large price swings i.e. investor can effectively use straddles when they expect a large move in a stock price but are uncertain about the direction \citep{hull2003options}. Effectively hedging the risk of any adverse price movements.  \cite{alexiou2021pricing} posits that investors pay a substantial premium for straddles to hedge the risk of increased volatility and large stock price on earnings announcements. 
\\
\begin{figure}[H]
  \begin{minipage}[b]{0.60\linewidth}
    \centering
      \includegraphics[height=2.7in, width = 3in, trim=6 0 70 5, clip]{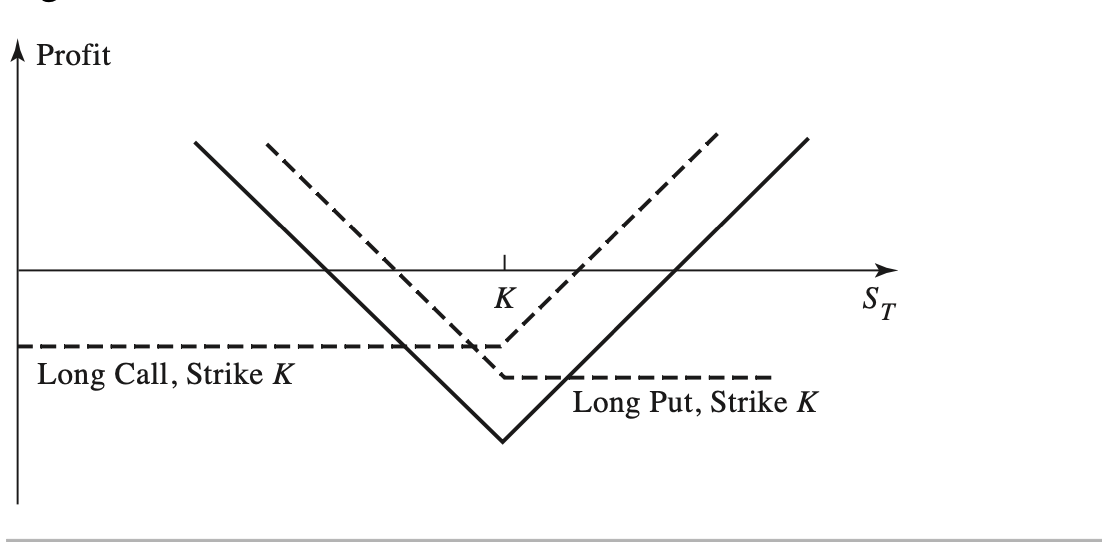}
    \par\vspace{0pt}
  \end{minipage}%
  \begin{minipage}[b]{0.2\linewidth}
    \centering%
  \includegraphics[height=2.7in, width = 3in, trim=0 -5 0 0, clip]{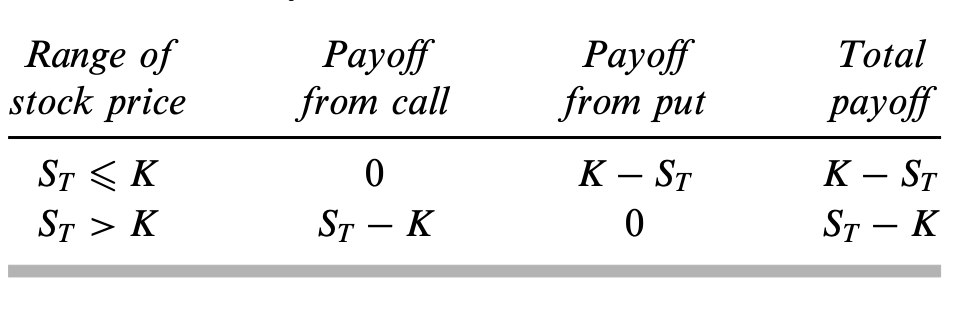}
    \par\vspace{0pt}
  \end{minipage}
\caption{Profit Diagram Straddle.\citep{hull2003options}. }
\label{fig:Straddle Profit}
\end{figure}

Assuming perfect market conditions we investigate this proposition by calculating the cost and profit of delta-neutral straddle before and after earning announcements whiles analyzing and comparing the outcomes of  concave and non-concave IV curves. \cite{alexiou2021pricing}
propose calculating the formation of the straddle by computing the following algorithm:
Using options, with the shortest time to maturity usually expiries within 3 to 14 days to create the delta neutral long call and short call. The returns are computed as:
 \begin{equation}\label{eq:straddlereturn}
    STRADDLE = wR_{c} + (1 - w)R_{p} 
    \end{equation}
 
  \begin{equation}\label{eq:weight}
    w = - \frac{\Delta_{p}/P}{\Delta_{c}/C - \Delta_{p}/P}
    \end{equation}
\\
Where $\Delta_{p}(\Delta_{c}) $ = Delta of a Put(Call) , P = Put Price ,  C = Call Price , $R_{p} (R_{p} )$ = Returns of Put/Call on EAD\\
w = wieght that ensure the straddle is delta neutral on formation.
It is known at the firm level that effect of this proposition is that typically profitable to write straddles to profit of the uncertainty observed. (see \cite{gao_xing_zhang_2018} and \cite{dubinsky2019option} ) ,  Further as in \cite{alexiou2021pricing} we compute  a measure of the percentage change in the stock price in either direction needed to offset the cost of a symmetric ATM straddle. \\

  \begin{equation}\label{eq:impmove}
     IMPMOVE = \frac{(C+P) }{S} 
    \end{equation}\\
  Where S = Current asset price at formation\\
    
Further we study the characteristics of the concave and non-concave IV curves to discern any commonality and major differences in both curves.\\

\subsection{Formation of Strangle and Implications on IV Curve}
Straddles are structured to benefit from shifts in prices greater than the ATM prices. When prices increase, the resulting change in option value is captured by delta neutral straddles. Price jumps violates this neutrality, if investors anticipate huge jumps in price movements  they can suitably use strangles to hedge the gamma risk. Strangles have a lower cost compared to straddles but, the downside is  a price change in either direction has to be sufficiently large for a strangle to pay off. The further  the strikes are apart, the less the downside risk and the farther the stock price has to move for a profit to be realized. \cite{hulland} . Purchasing an out the money (OTM) call and a put, investors bet there will be a large price move in either direction to the current price. We imply that in the presence of concave implied volatilities prices are very volatile and investors are willing to pay substantial premiums to hedge the risk surrounding earnings.  \cite{alexiou2021pricing} study provides evidence to this statement. Graphical representation of the profit and payoff below use the same example above:
\begin{figure}[H]
  \begin{minipage}[b]{0.60\linewidth}
    \centering
      \includegraphics[height=2.7in, width = 3in, trim=6 0 70 5, clip]{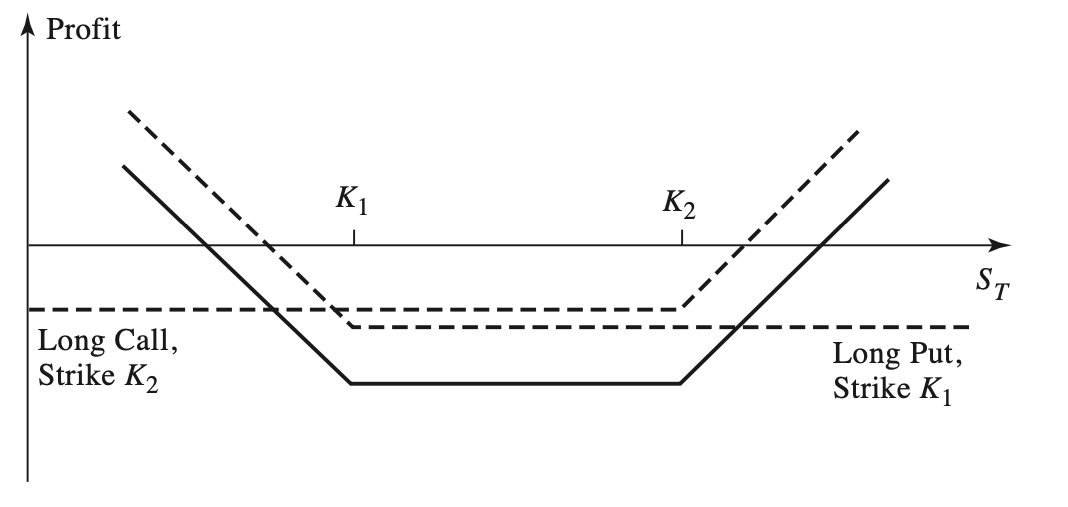}
    \par\vspace{0pt}
  \end{minipage}%
  \begin{minipage}[b]{0.2\linewidth}
    \centering%
  \includegraphics[height=2.7in, width = 3in, trim=0 -5 0 0, clip]{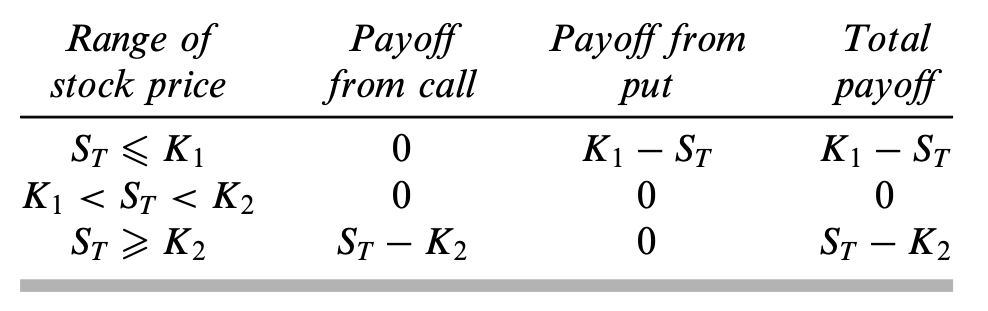}
    \par\vspace{0pt}
  \end{minipage}
\caption{Profit Diagram Strangle.\citep{hull2003options}. }
\label{fig:Strangle Profit}
\end{figure} 

\subsection{Characteristics of Option and Price Data for Quarterly Earnings Announcement (2021 -2022) }
To illustrate the above statements in relation to earning announcement returns for strangle and straddle, an analysis of 49 quarterly earning announcements over a span of 2021- 2022 is presented in this paper (see. Appendix* for names).  The average (median - max) implied volatility for ATM options before the earning announcement reports 75.70\% (71.60\%- 188.77\%) on the announcement the implied volatility observed respectively are 69.64\% (50\% - 235\%). For the average and median results, a drop in implied volatility indicates that news on the earning announcement is within market expectation but for the maximum values the huge spike reflects sudden news market did not anticipate. For our study period and data set the average (median - max) of price returns(logarithmic*) report -0.1\% (0.158\% - 7.03\%) (see \ref{tab:TableOptionCharac}). \\
(Table will be kept and/or moved to appendix)\\

\begin{table}[H]
\caption{Option Characteristics}
\label{tab:TableOptionCharac1}
\centering
\begin{tabular}{cccc}
  \toprule
   & Price Return & Before EAD Implied Vol ATM& EAD Implied Vol ATM \\
  \midrule 
  mean & -0.001053	 & 0.756912	 & 0.696365 \\
  \midrule
  min& -0.073728 & 0.2260 & 0.141604\\
  \midrule
  25\% & -0.022073& 0.42950 & 0.287501 \\
  \midrule
  50\% & 0.001580 & 0.7162& 0.5004 \\
  \midrule
  75\% & 0.020028 & 0.8935 & 0.8161 \\
  \midrule
  max& 0.070315 & 1.8877 & 2.3258 \\
  \bottomrule
\end{tabular}
\end{table}

\subsection{Characteristics of Concave vs Non-Concave IV for Quarterly Earnings Announcement (2021 -2022)}
We filter concave IV curves based on the step difference (0.025) , the  moneyness difference used in the Jackwerth implied probability estimator (see equation *) and an approximation to convexity (\ref{eq:concave}).  Twenty out of the sample exhibited concavity and from our estimation of the implied volatility curve (see Appendix or Reference) visibly show the concave smiles. (see. Appendix) \\
  \begin{equation}\label{eq:concave}
     CONVEXEST = Implied-Volatility(ATM - x\%) 
     + Implied-Volatility(ATM +x\%) 
         \end{equation}
      \[ - 2(Implied-Volatility ATM)\]

This section discusses the outcome of delta neutral strangle and straddles, we construct these strategies using options1-day before the earning announcements and closing the opened positions at the end of the trading day after the announcement. We compare selected samples that show concavity in their implied volatility curves a-day before the earning announcements to those without. Straddles are formed from long call and put ATM options, to generate profit the price movement must be large to offset the formation costs. Based on the impmove cost (\ref{eq:impmove}), an average(median) of 9.44\% (8.13\%) is needed to offset straddle cost in the presence of concave IV curves. Contrasting with 4.28\% (3.74\%) for straddle with non-concave IV curves, a significant 5.16\% (4.39\%) difference is observed suggesting that investors pay more to form straddles with concave IV curves. (see  \ref{tab:TableOptionCharac} for quartile). \\
Establishing this characteristic, we contrast price returns to the implied stock price move (\ref{eq:impmove}). Assets with concave IV have an average(median) return of -0.068\% (0.339\%) a difference of 9.37\% (7.79\%) to impmove this indicates that asset prices on average do not move significantly to offset the purchase cost of the straddle. For non-concave IV curves -0.13\%(0.158\%) are the  average(median) return with a difference of 4.15\%( 3.58\%). We infer investors purchasing concave IV straddles are more concerned about the uncertainty surrounding the earnings announcement and are willing to accommodate higher premiums to hedge against unfavorable market moves. 
\begin{table}[H]
\caption{Price Return vs IMPMOVE Concave \& Non-Concave Straddles}
\label{tab:PvsReturn}
\centering
\begin{tabular}{cccc}
    \hline
    \multicolumn{2}{c}{Concave Straddle} \\%
     \hline
  \toprule
   & IMPMOVE&  Price Return & Difference \\
  \midrule 
  mean & 0.094425	 &-0.000682 & 0.093744 \\
  \midrule
  50\% & 0.081304 & 0.003388& 0.077916 \\
  \midrule
  max& 0.189257 & 0.063732 & 0.125525 \\
    \hline
    \multicolumn{2}{c}{Non-Concave Straddle} \\%
  \hline
  \toprule
   & IMPMOVE&  Price Return & Difference \\
  \midrule 
  mean & 0.042823	 &-0.000682	 & 0.041514\\
  \midrule
  50\% & 0.037380 & 0.003388& 0.035799 \\
  \midrule
  max& 0.079565 & 0.063732 & 0.009249 \\
  
  \bottomrule
\end{tabular}
\end{table}

Appendix(\ref{fig:tabstrad}) for Tables for all assets and option strategies for 7 quarterly earnings announcements. \\

We construct strangles by taking strike prices 50 basis points from the ATM money price in both directions. Moneyness levels are based on the Jackwerth  (section/part/table ref ) algorithm, i.e a constant moneyness difference of 25 basis points. For strangles we designed the put and call such that price must change two moneyness levels in either direction of the ATM price for a profit to be realized. This then takes into consideration the gamma effect in the underlying prices. For the samples (reference the data set*) indicated we investigate the cost and relate it to price return. The average(median) impmove for strangles is reported 
at 5.32\%(4\%) these values are far lower than straddle ATM impmove as a result of a lower probability of payoff. Comparing to 1.27\%(0.86\%) for non-concave IV curves the cost difference becomes more noticeable(see Table \ref{tab:TableOptionCharac2}). \\
Further contrasting price returns to impmove concave IV , an average(median) of -0.0682\%(0.339\%) with a difference of 5.26\%( 3.65\%) indicates that investors will pay more to hedge the uncertainty of price direction on earnings announcement. Non concave IV present higher price returns and lower difference with impmove Further substantiating the assertion that options are more costly with the presence of concave IV.(see. Table .\ref{tab:PvsReturn1})\\
We have a further look at the returns of strangles and straddles, the result show the average(median) return for straddles are below strangles(see stats . This is a result of the high cost of straddles, returns are setoff by the cost at formation (see Appendix\ref{fig:Strangle G Profit}). 

\begin{figure}[H]
  \begin{minipage}[b]{0.60\linewidth}
    \centering

      \includegraphics[height=2.7in, width = 3in, trim=6 0 70 5, clip]{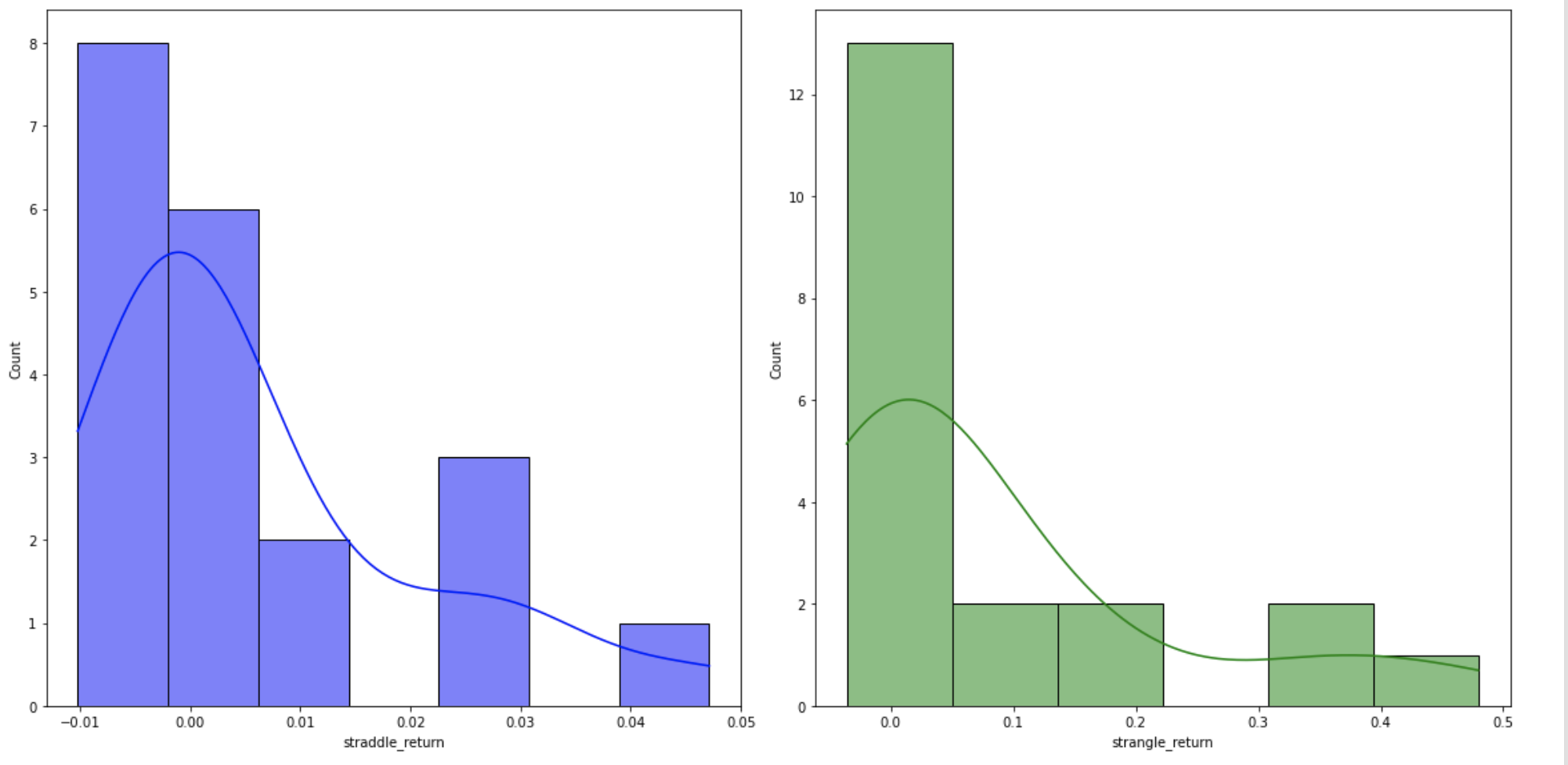}
      
    \par\vspace{0pt}
  \end{minipage}%
  \begin{minipage}[b]{0.2\linewidth}
   \centering%
  \includegraphics[height=2.7in, width = 3in, trim=0 -5 0 0, clip]{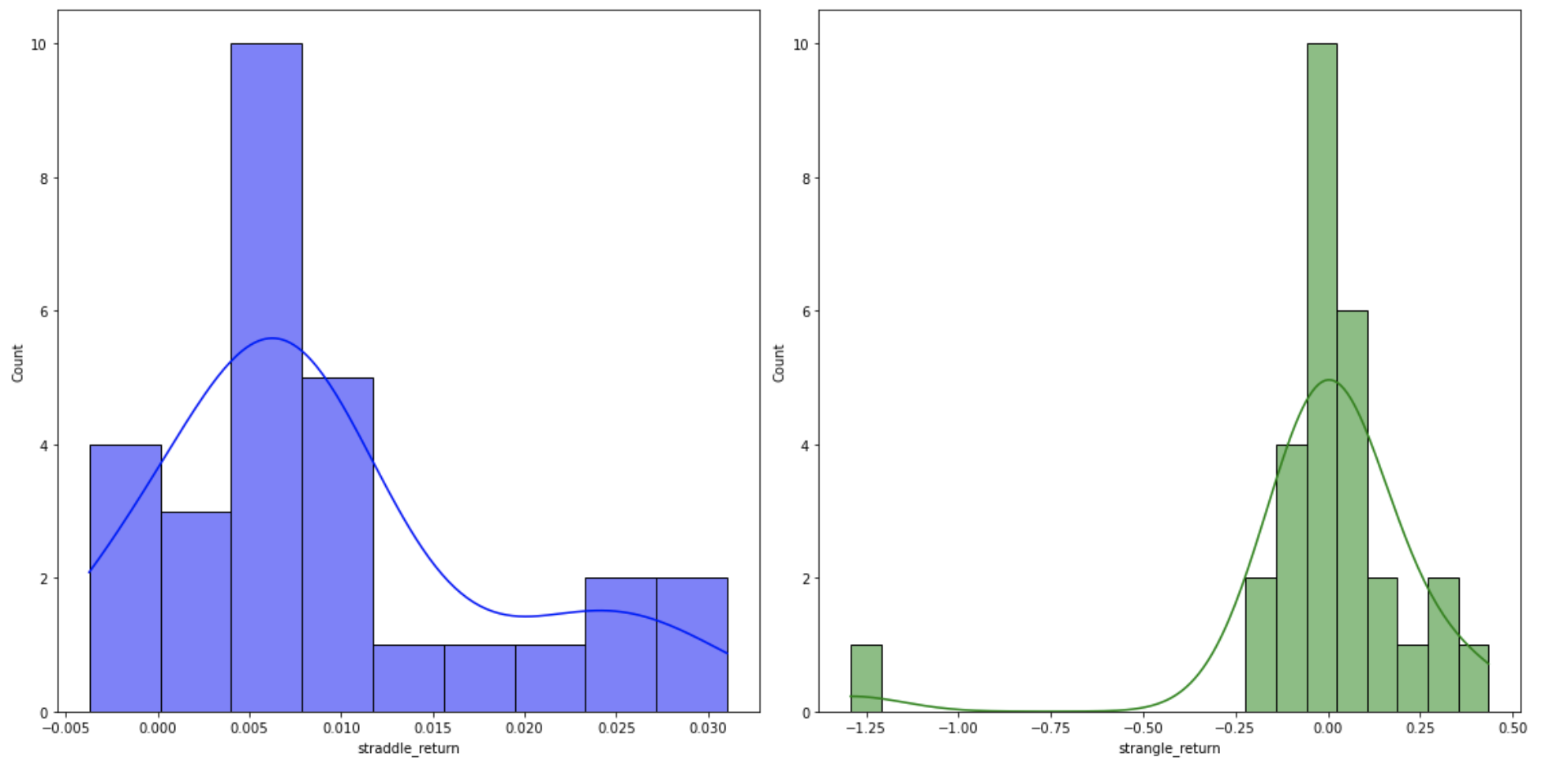}
    \par\vspace{0pt}
  \end{minipage}
\caption{Histogram of Straddle vs Strangle, Concave(Left) ,Non-concave (Right) }
\label{fig:hist}
\end{figure}

Table( \ref{fig:Strangle G Profit})below details the returns in both concave and nonconcave IV curves. Particularly noticeable are the higher  average strangle returns for concave and non concave IV. As indicated in previous sections high volatility is a  feature of concave IV  curves , this feature gives the underlying asset momentum to exceed the strike prices at formation. Juxtaposed , non-concave IV have lesser volatility , prices do not rise sufficiently enough to offset the cost hence negative returns when the position is closed. \\

\section{Conclusion}\label{sec:conclude}
We show that the IV smiles of stock options commonly have concavity before the EAD. This form contrasts sharply with the convex volatility smiles often observed with equities options. Concavity is especially noticeable in short-term options; it shows a bimodal risk-neutral distribution for the underlying stock price and disappears after the announcement as the uncertainty around this event is resolved. Overall, we show that investors can predict which announcements will cause larger-than-average stock price swings and will pay a significant premium to protect against this event risk. This hedging behaviour affects options pricing, resulting in a concave IV curve. We find that concavity in the IV curve is an ex-ante option-implied signal for event risk in the underlying stock due to the upcoming announcement. \\
Our research focuses on scheduled company earnings presentations. However, it would be fascinating to investigate the characteristics and informative quality of IV curves around other non-corporate events that may cause substantial asset price movements. Prior research has claimed that macroeconomic announcements and geopolitical events can create significant risk, which can be reflected in options prices. We predict that the curvature qualities of the IV curve surrounding these events will offer significant information about event risk pricing and asset price behaviour. The investigation of these impacts is left to future research.\\

\newpage

\begin{appendix}

\begin{table}[H]
\caption{IMPMOVE Concave \& Non-Concave Straddles}
\label{tab:TableOptionCharac}
\centering
\begin{tabular}{cccc}
  \toprule
   & Concave &  Non-Concave& Difference \\
  \midrule 
  mean & 0.094425	 &0.042823	 & 0.051603 \\
  \midrule
  min& 0.048125 &0.020311&0.027814\\
  \midrule
  25\% & 0.064146 & 0.031975 & 0.032171 \\
  \midrule
  50\% & 0.081304 & 0.037380& 0.043924 \\
  \midrule
  75\% & 0.126696 & 0.052287 & 0.074409 \\
  \midrule
  max& 0.189257 & 0.079565 & 0.109693 \\
  \bottomrule
\end{tabular}
\end{table}

\begin{table}[H]
\caption{IMPMOVE Concave \& Non-Concave Strangles}
\label{tab:TableOptionCharac2}
\centering
\begin{tabular}{cccc}
  \toprule
   & Concave &  Non-Concave& Difference \\
  \midrule 
  mean & 0.053273	 &0.012713	 & 0.040560 \\
  \midrule
  min& 0.013878 &0.002798&0.011080\\
  \midrule
  25\% & 0.025122 & 0.005641 & 0.019480 \\
  \midrule
  50\% & 0.039953 & 0.008552& 0.031400 \\
  \midrule
  75\% & 0.082322& 0.016902& 0.065420 \\
  \midrule
  max&0.142047 & 0.039461 & 0.102585 \\
  \bottomrule
\end{tabular}
\end{table}

\begin{table}[H]
\caption{Price Return vs IMPMOVE Concave \& Non-Concave Straddles}
\label{tab:PvsReturn1}
\centering
\begin{tabular}{cccc}
    \hline
    \multicolumn{2}{c}{Concave Straddle} \\%
     \hline
  \toprule
   & IMPMOVE&  Price Return & Difference \\
  \midrule 
  mean & 0.094425	 &-0.000682 & 0.093744 \\
  \midrule
  min& 0.048125 &-0.073728& 0.025603\\
  \midrule
  25\% & 0.064146 & -0.027575 & 0.032171 \\
  \midrule
  50\% & 0.081304 & 0.003388& 0.077916 \\
  \midrule
  75\% & 0.126696 & 0.025162	 & 0.101534 \\
  \midrule
  max& 0.189257 & 0.063732 & 0.125525 \\
    \hline
    \multicolumn{2}{c}{Non-Concave Straddle} \\%
  \hline
  \toprule
   & IMPMOVE&  Price Return & Difference \\
  \midrule 
  mean & 0.042823	 &-0.000682	 & 0.041514\\
  \midrule
  min& 0.020311 &-0.073728& 0.050903\\
  \midrule
  25\% & 0.031975 & -0.027575 & 0.012198\\
  \midrule
  50\% & 0.037380 & 0.003388& 0.035799 \\
  \midrule
  75\% & 0.052287 & 0.025162	 & 0.032992 \\
  \midrule
  max& 0.079565 & 0.063732 & 0.009249 \\

  \bottomrule
\end{tabular}
\end{table}

\begin{table}[H]
\caption{Price Return vs IMPMOVE Concave \& Non-Concave Strangles}
\label{tab:PvsReturn3}
\centering
\begin{tabular}{cccc}
    \hline
    \multicolumn{2}{c}{Concave Strangle} \\%
     \hline
  \toprule
   & IMPMOVE&  Price Return & Difference \\
  \midrule 
  mean &0.053273	 &-0.000682& 0.052591 \\
  \midrule
  min& 0.013878 &-0.073728& 0.059850\\
  \midrule
  25\% & 0.025122 & -0.027575 & 0.002453 \\
  \midrule
  50\% &0.003388& 0.001580& 0.036565\\
  \midrule
  75\% & 0.025162& 0.019294	 & 0.057161\\
  \midrule
  max& 0.063732& 0.070315 & 0.078315\\
    \hline
    \multicolumn{2}{c}{Non-Concave Strangle} \\%
  \hline
  \toprule
   & IMPMOVE&  Price Return & Difference \\
  \midrule 
  mean & 0.012713	 &-0.001309	 & 0.051964\\
  \midrule
  min& 0.002798 &-0.071214& 0.057337\\
  \midrule
  25\% & 0.005641& -0.019777 & 0.005345\\
  \midrule
  50\% & 0.008552 & 0.001580& 0.038373\\
  \midrule
  75\% & 0.016902& 0.019294	 & 0.063028 \\
  \midrule
  max& 0.039461 & 0.070315& 0.071732 \\

  \bottomrule
\end{tabular}
\end{table}

\begin{figure}[H]
  \begin{minipage}[b]{0.50\linewidth}
    \centering
    
\begin{table}[H]
\label{tab:PvsReturn5}
\centering
\begin{tabular}{cccc}
    \hline
    \multicolumn{2}{c}{Concave Returns} \\%
     \hline
  \toprule
   & Straddle&  Strangle  & Difference \\
  \midrule 
  mean & 0.006285	 &0.084590	 & 0.078305\\
  \midrule
  min& -0.010214&-0.035842& 0.025627\\
  \midrule
  25\% & -0.002387& 0.003203 & 0.000816\\
  \midrule
  50\% & -0.000357 & 0.020802& 0.020445\\
  \midrule
  75\% & 0.009355& 0.124716	 & 0.115361\\
  \midrule
  max& 0.047136 & 0.479707& 0.432572 \\
  
  \bottomrule
\end{tabular}
\end{table}

    \par\vspace{0pt}
  \end{minipage}%
  \begin{minipage}[b]{0.5\linewidth}
    \centering%
    
\begin{table}[H]
\label{tab:PvsReturn6}
\centering
\begin{tabular}{cccc}
    \hline
    \multicolumn{2}{c}{Non - Concave Returns} \\%
     \hline
  \toprule
   & Straddle&  Strangle  & Difference \\
  \midrule 
  mean & 0.009214	 &-0.008804	 & 0.000410\\
  \midrule
  min& -0.003720&-1.292246& 1.288526\\
  \midrule
  25\% & 0.005181& -0.058201& 0.053020\\
  \midrule
  50\% & 0.006681 & 0.005598& 0.001083\\
  \midrule
  75\% & 0.011587& 0.068349	 & 0.056762\\
  \midrule
  max& 0.031038 & 0.435409& 0.404371 \\

  \bottomrule
\end{tabular}

\end{table}

    \par\vspace{0pt}
  \end{minipage}
\caption{ Straddle \& Strangle  Returns }
\label{fig:Strangle G Profit}
\end{figure}



\begin{figure}[H]
    \centering
 \includegraphics[height=7in, width = 7in, trim=0 0 0 0, clip]{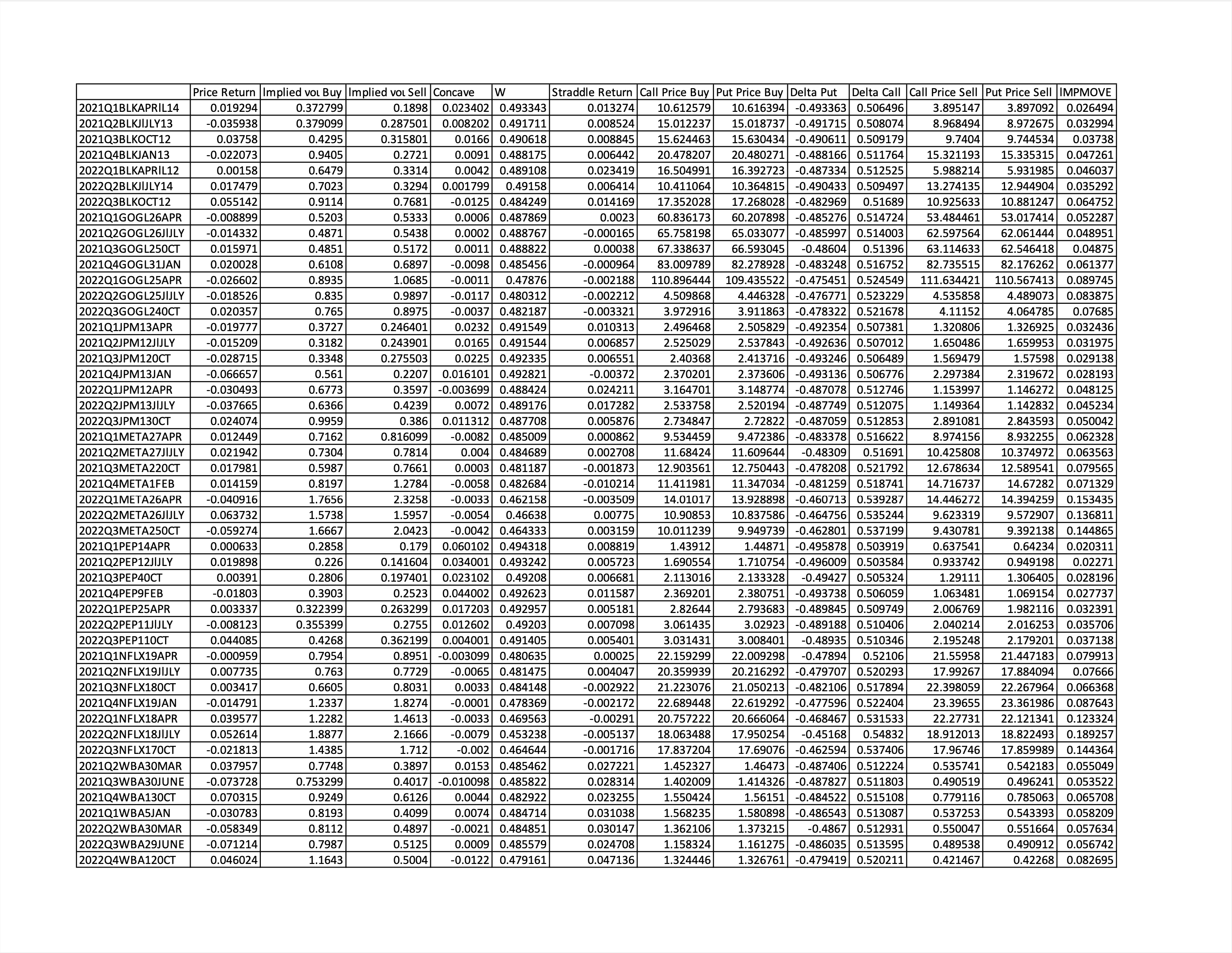}
 \caption{All Companies and Quarters Straddle Data} 
 \label{fig:tabstrad}
\end{figure}

\begin{figure}[H]
    \centering
 \includegraphics[height=7in, width = 7in, trim=0 0 0 0, clip]{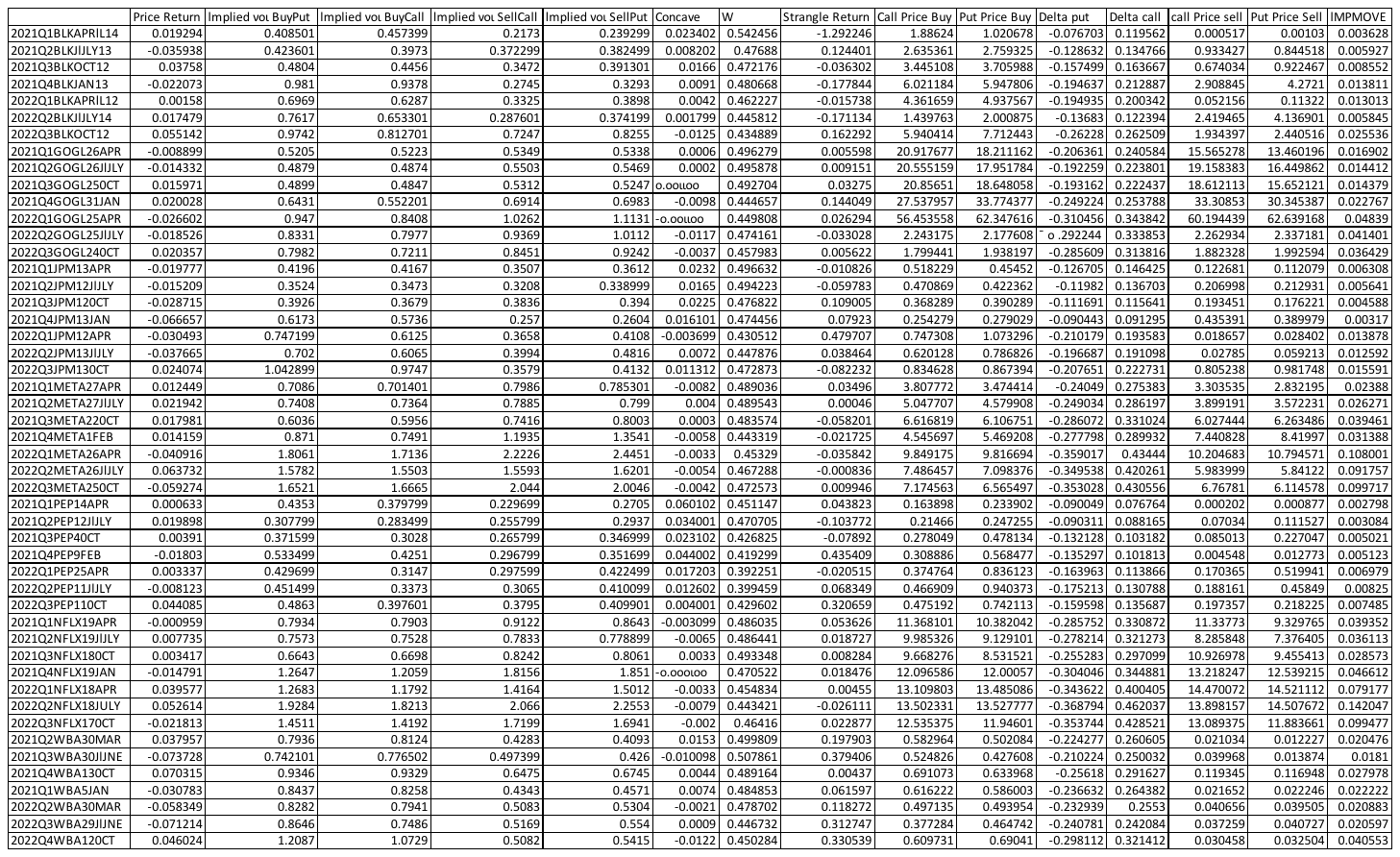}
 \caption{All Companies and Quarters Strangle Data} 
 \label{fig:nflx1}
\end{figure}

\newpage

\section{APPENDIX A - Fast and Stable Method}\label{sec:A}
 \begin{center}
 
 To implement the method, we mechanically make a series of moneyness values separated equally, i.e. by $\Delta $= 0.025 and $\lambda$ =0.01\footnote{The same values is taken in Jackwerth, Jens Carsten, and Mark Rubinstein. Recovering probabilities and risk aversion from option prices and realized returns. No. 11638. University Library of Munich, Germany, 2004.}. These include the ones we already have IV values taken from Bloomberg and the ones we need to calculate to form the IV curve. For our project, the moneyness values were common for all curves; they were,\\
 
 \begin{equation}\label{eq: m}
    M_{j} = M_{0} + j*\Delta
  \end{equation}
  $M_{j}$ = New Moneyness value\\
$M_{0}$= Lowest Moneyness value \\
$\Delta $= difference between two adjacent moneyness values\\
j =0,….,16\\

To find the implied volatility for strikes/moneyness that we don’t directly see in the market, we minimize the following objective function:\\
  \begin{equation}\label{eq:jack1}
 min(\sigma_{j})\frac{\Delta^{4}}{2(J+1)}\sum_{j=0}^{j}(\sigma_{j}^{''} )^{2} +  \frac{\lambda}{2I} \sum_{i=0}^{I}(\sigma_{i}  - \sigma_{i}^{-}  )^{2}
    \end{equation}\\

$\sigma_{j} $=implied volatility associated with strike price$ k_{j}$\\
$\Delta $ = difference between two adjacent future index values\\
$ \lambda $= trade-off parameter for balancing smoothness versus fit\\
$\sigma_{j}^{''}$= second derivative of implied volatility concerning strike prices, numerically approximate.\\
$\sigma_{i}$= implied volatility associated with strike price$ k_{i}$\\
$\sigma_{i}^{-}$=observed implied volatility of option with strike price $k_{i}$\\

The first-order conditions for index j when there is no observed IV value are:\\
 \begin{equation}\label{eq: m}
    \sigma_{j-2} - 4\sigma_{j-1} + 6\sigma_{j} - 4\sigma_{j+1} + \sigma_{j+2} = 0
  \end{equation}
  For j = 0, 1, J – 1, and J, we set the missing implied volatilities outside the range of j to $\sigma_{0} $and $\sigma{J}. $
  When there is an observed IV value for index j, the first-order conditions are:\\

 \begin{equation}\label{eq: m}
    \sigma_{j-2} - 4\sigma_{j-1} +   [6 + (\lambda_(J+1)/(I\Delta^{4}))]\sigma_{j}   -   4\sigma_{j+1} + \sigma_{j+2} = [6 + (\lambda(J+1)/(I\Delta^{4}))]\sigma_{j}^{-}
  \end{equation}

$\sigma_{j}^{-}$= observed implied volatility of option with moneyness value $M_{j}$

\end{center}
Once we have solved the system of first-order conditions for the optimal implied volatility function, we calculate the Black–Scholes \citep{black1973pricing} options prices to approximate the risk-neutral distribution, using the Finite difference method. The risk-neutral distribution is given as  $G(k)= e^{rT}(d^{2}c/dk^{2})$. The numerical approximation is :\\

  \begin{equation}\label{eq:gk}
G(k)\approx e^{rt}(c_{1} + c_{3} - 2c_{2}) /\Delta^{2}
    \end{equation}\\

\end{appendix}

\bibliography{bibliografi}

 
\end{document}